\documentclass[trackchanges,twocolumn]{aastex7}
\usepackage[version=4]{mhchem}
\usepackage{upgreek}
\usepackage{CJK}

\begin{document}
\begin{CJK*}{UTF8}{gbsn}

\title{JWST/MIRI Hydrocarbon and Water Absorption in the Wind of a Young Disk: Signatures of Pebble Drift and Carbon Grain Sublimation}

\author[orcid=0000-0002-5296-6232,gname='Maria Jose']{Mar\'ia Jos\'e Colmenares}
\affiliation{Department of Astronomy, University of Michigan, 1085 South University Avenue, Ann Arbor, MI 48109, USA}
\email[show]{mjcolmen@umich.edu}

\author[orcid=0000-0003-4179-6394,gname='Edwin A.']{Edwin A. Bergin} 
\affiliation{Department of Astronomy, University of Michigan, 1085 South University Avenue, Ann Arbor, MI 48109, USA}
\email{ebergin@umich.edu}

\author[orcid=0000-0002-0661-7517]{Ke Zhang} 
\affiliation{Department of Astronomy, University of Wisconsin-Madison, 475 N Charter St, Madison, WI 53706, USA}
\email{ke.zhang@wisc.edu}

\author[0000-0003-0787-1610,gname='Geoffrey A.']{Geoffrey A. Blake}
\affiliation{Division of Geological and Planetary Sciences, California Institute of Technology, MC 150-21, 1200 E California Boulevard, Pasadena, CA 91125, USA}
\email{}

\author[0000-0001-7552-1562, gname='Klaus M.']{Klaus M. Pontoppidan}
\affiliation{Jet Propulsion Laboratory, California Institute of Technology, 4800 Oak Grove Drive, Pasadena, CA 91109, USA}
\affiliation{Division of Geological and Planetary Sciences, California Institute of Technology, MC 150-21, 1200 E California Boulevard, Pasadena, CA 91125, USA}
\email{}

\author[0000-0002-4876-630X, gname='Alexa R.']{Alexa R. Anderson}
\affiliation{Institute for Astronomy, University of Hawai`i at Mānoa, 2680 Woodlawn Drive, Honolulu, HI 96822, USA}
\email{alexaand@hawaii.edu}

\author[0000-0002-6695-3977]{John Carr}
\affiliation{Department of Astronomy, University of Maryland, College Park, MD 20742, USA}
\email{}

\author[0000-0003-2985-1514]{Emma Dahl}
\affiliation{California Institute of Technology, 4800 Oak Grove Drive, Pasadena, CA 91109, USA}
\email{}

\author[0000-0002-5758-150X]{Joan Najita}
\affiliation{NSFs NOIRLab, 950 N. Cherry Avenue, Tucson, AZ 85719, USA}
\email{}

\author[0000-0001-5058-695X,gname='Jonathan P.']{Jonathan P. Williams}
\affiliation{Institute for Astronomy, University of Hawai`i at M\={a}noa, 2680 Woodlawn Drive, Honolulu, HI 96822, USA}
\email{}

\author[0000-0003-3682-6632]{Colette Salyk}
\affiliation{Vassar College, 124 Raymond Avenue, Poughkeepsie, NY 12604, USA}
\email{}

\author[0000-0001-8240-978X]{Till Kaeufer}
\affiliation{Department of Physics and Astronomy, University of Exeter, Exeter EX4 4QL, UK}
\email{}

\author[0000-0002-0554-1151]{Mayank Narang}
\affiliation{Jet Propulsion Laboratory, California Institute of Technology, 4800 Oak Grove Drive, Pasadena, CA 91109, USA}
\email{}

\author[0000-0001-7962-1683]{Ilaria Pascucci}
\affiliation{Lunar and Planetary Laboratory, University of Arizona, Tucson, AZ 85721, USA}
\email{}

\author[0000-0002-1103-3225]{Beno\^it Tabone}
\affiliation{Universit\'{e} Paris-Saclay, CNRS, Institut d’Astrophysique Spatiale, F-91405 Orsay, France}
\email{}

\author[0000-0002-2828-1153]{Lucas Cieza}
\affiliation{Instituto de Estudios Astrof\'isicos, Universidad Diego Portales, Av. Ejercito 441, Santiago, Chile}
\affiliation{Millennium Nucleus on Young Exoplanets and their Moons (YEMS), Santiago, Chile}
\email{}

\author[0000-0002-4147-3846]{Miguel Vioque}
\affiliation{European Southern Observatory, Karl-Schwarzschild-Str. 2, 85748 Garching bei M\"{u}nchen, Germany}
\email{miguel.vioque@eso.org}

\author[0000-0001-8790-9011]{Adrien Houge} 
\affiliation{Center for Star and Planet Formation, GLOBE Institute, University of Copenhagen, Øster Voldgade 5-7, 1350 Copenhagen,Denmark}
\email{}

\author[0000-0002-3291-6887]{Sebastiaan Krijt}
\affiliation{ School of Physics and Astronomy, University of Exeter, Stocker Road, Exeter, EX4 4QL, UK}
\email{}

\author[0000-0001-8407-4020,gname='Aditya M.']{Aditya M. Arabhavi}
\affiliation{Jet Propulsion Laboratory, California Institute of Technology, 4800 Oak Grove Drive, Pasadena, CA 91109, USA}
\email{}

\author[0000-0003-4853-5736]{Giovanni Rosotti}
\affiliation{Dipartimento di Fisica,Universit\`{a} degli Studidi Milano,Via Celoria 16,I-20133 Milano, Italy}
\email{giovanni.rosotti@unimi.it}

\author[0000-0003-2251-0602]{John Carpenter}
\affiliation{Joint ALMA Observatory, Alonso de C\'ordova 3107, Vitacura, Santiago 763-0355, Chile}
\email{john.carpenter@alma.cl}

\author[0000-0002-7607-719X]{Feng Long (龙凤)}
\affiliation{Kavli Institute for Astronomy and Astrophysics, Peking University, Beijing 100871, China}
\email{long.feng@pku.edu.cn}

\author[0000-0001-8764-1780]{Paola Pinilla}
\affiliation{Mullard Space Science Laboratory, University College London, Holmbury St Mary, Dorking, Surrey RH5 6NT, UK}
\email{}

\author[0000-0003-0386-2178]{Jayatee Kanwar}
\affiliation{Department of Astronomy, University of Michigan, 1085 South University Avenue, Ann Arbor, MI 48109, USA}
\email{}

\author[0009-0002-2380-6683]{Eshan Raul} 
\affiliation{Department of Astronomy, University of Wisconsin-Madison, 475 N Charter St, Madison, WI 53706, USA}
\email{}

\author[0000-0001-8284-4343]{Karina Mauco}
\affiliation{Universidad Nacional Aut\'onoma de M\'exico, Instituto de Astronom\'ia, AP 106, Ensenada 22800, BC, M\'exico}
\email{}

\author[0000-0002-1575-680X]{James Miley}
\affiliation{Departamento de F\'isica, Universidad de Santiago de Chile, Avenida Victor Jara 3659, Santiago, Chile}
\email{}

\author[0000-0002-1566-389X]{Abygail Waggoner}
\affiliation{Department of Astronomy, University of Wisconsin-Madison, 475 N Charter St, Madison, WI 53706, USA}
\email{}

\collaboration{all}{and the JDISCS collaboration}

\received{February 16th, 2026}

\accepted{April 9th, 2026}

\begin{abstract}

We present JWST/MIRI-MRS observations of ISO-Oph~37, a highly inclined flat-spectrum ($\lesssim$1\,Myr old) source, to investigate the chemical composition and dynamical origin of its inner-disk gas. The spectrum reveals a rich combination of molecular emission and absorption: \ce{H2O}, CO, and OH are detected in emission, while strong absorption is observed from CO, \ce{H2O}, \ce{CO2}, HCN, \ce{C2H2}, and \ce{CH4}, with no detectable ice absorption features. LTE slab modeling of the absorption yields excitation temperatures of $T_{\rm ex}\sim400$-600~K and column densities of $\log N/{\rm cm}^{2}\sim16$-19, characteristic of warm gas located within the inner few au.
The absorption lines are significantly blueshifted relative to the systemic velocity, with mid-IR lines exhibiting larger shifts than near-IR CO absorption. This velocity structure points to a velocity- and temperature-stratified molecular disk wind. In this framework, the absorption directly samples disk material lifted from the inner disk surface, preserving the chemical imprint of the wind-launching region.
Along the line of sight, ISO-Oph~37 is unusually hydrocarbon-rich compared to other known absorption systems (GV~Tau~N and IRS~46), exhibiting high (\ce{C2H2}+\ce{CH4})/HCN, (\ce{C2H2}+\ce{CH4})/CO and \ce{H2O}/CO column density ratios, while the CO and HCN columns remain broadly typical. We find that these molecular ratios are best explained by enhancement of both hydrocarbons and water, driven by inward drift and sublimation of icy pebbles and by thermal processing of carbonaceous grains at the soot line. ISO-Oph~37 thus demonstrates that carbon-rich inner-disk chemistry can be established early in disk evolution and that it can be directly probed through molecular absorption in disk winds.

\end{abstract}

\keywords{\uat{Protoplanetary disks}{1300}  --- \uat{Protostars}{1302} ---  \uat{Planet formation}{1241} --- \uat{Circumstellar disks}{235} --- \uat{Astrochemistry}{75}}

\section{Introduction} 

Understanding the carbon budget in the inner regions of protoplanetary disks is central to piecing together the formation of terrestrial planets and the chemistry from which they emerge \citep{madhusudhan2012,Oberg21, bergin2024_coratios}. Carbon-bearing species ranging from simple volatiles like \ce{CO2} and \ce{CH4} to larger hydrocarbons such as \ce{C2H2} play pivotal roles in disk chemistry,  influencing condensable carbon carriers, and the initial conditions for planetary atmospheres, as well as the bulk carbon content of terrestrial planets. The unprecedented mid-infrared sensitivity and spectral resolution of the James Webb Space Telescope (JWST) \citep{rigby2023} has recently unveiled a rich inventory of hydrocarbons in inner disks, including \ce{CH4}, \ce{C2H2}, and even larger carbon chains, underscoring a carbon-rich chemistry with C/O ratios exceeding unity in some systems \citep{tabone2023,arabhavi2024, kanwar2024,kanwar2025b,colmenares2024b}. These discoveries highlight the presence of substantial carbon reservoirs in the warm inner disk.

Prior to JWST, much of our understanding of inner disk chemistry came from \textit{Spitzer}/Infrared Spectrograph (IRS) surveys, which revealed numerous molecules such as \ce{H2O}, \ce{HCN}, \ce{C2H2}, OH, and \ce{CO2}, predominantly in emission from the warm surface layers of disks within a few astronomical units \citep{Carr11, Pascucci2009, pontoppidan2010,salyk2011,Pascucci13}. These observations established that disks are chemically active environments where both oxygen- and carbon-bearing molecules are abundant. An emerging trend has suggested that as a given disk evolves, the inner gas becomes more carbon-rich, in line with transport-chemistry models in which icy pebble drift and trapping beyond the \ce{H2O} and \ce{CO2} snowlines sequester oxygen while grain-surface processing and inward gas flows can deliver carbon-bearing volatiles (including \ce{CH4}) to the inner disk, raising the gas-phase C/O ratio \citep{Mah23, mah2024,sellek2025}. Early JWST-MIRI spectra provide supporting evidence, with some disks at later stages showing enhanced hydrocarbons and elevated C/O \citep{arabhavi2025,jang2025,long2025}.

However, the chemical makeup of protoplanetary disks has not always revealed itself through molecular emission lines. A small but important subset of \textit{Spitzer} targets, most notably the nearly edge-on systems IRS 46 and GV Tau, exhibit strong molecular absorption features \citep{lahuis2006,bast2013}. In these sources, \ce{C2H2}, \ce{CO2}, and \ce{HCN} are detected in absorption against the warm continuum of the inner disk, with excitation temperatures of $\sim$400-700\,K for \ce{C2H2} and \ce{HCN} and $\sim$250\,K for \ce{CO2}. Absorption offers a complementary and often more direct constraint on column densities along the line of sight to the inner continuum, reducing degeneracies tied to emitting area that can affect emission analyses, and thereby helping to localize warm gas vertically and radially in the inner disk \citep[e.g.,][]{bast2013}. The interpretation of these rare absorption sources has centered on their geometry since they are highly inclined disks that place warm gas in front of the bright continuum \citep{gibb2013}, but additional scenarios such as disk winds or accretion inflows have also been invoked \citep{lahuis2006,doppmann2008,najita2021}.

With JWST, absorption is now more easily identified; surveys of protostars and edge-on systems report mid-IR absorption (ices and, in some cases, gas) coexisting with line emission. These observations expand the limited \textit{Spitzer} sample and reframing absorption as both a geometric and evolutionary diagnostic rather than a rarity \citep{brunken2024,vangelder2024,tyagi2025,mcclure2025}. In such datasets, H$_2$ rotational lines and forbidden lines (e.g., [Ne\,\textsc{ii}], [Ne\,\textsc{iii}], [Fe\,\textsc{ii}], [Ar\,\textsc{ii/iii}]) can be spatially extended, enabling separation of disk, envelope, and wind/jet contributions relative to the continuum \citep[e.g.,][M. Narang et al. 2026, submitted]{bajaj2024,arulanantham2024,pascucci2025,narang2025}.

A persistent difficulty in studying most molecular absorption features is the disentanglement of the disk from its environment. Many systems still retain residual envelopes from the star formation process, or are located in regions of high extinction, such as $\rho$~Ophiuchi. This complicates the determination of their ages and evolutionary states, as envelope emission, scattered light, or foreground absorption can contaminate the disk signatures \citep{McClure10, sturm2024}. Moreover, the mid-infrared extinction law in $\rho$~Ophiuchi departs from diffuse-ISM prescriptions and varies with column density, which can alter continuum slopes and bias spectral energy distribution (SED) classifications, complicating the separation of truly embedded sources from edge-on Class II disks \citep{McClure2009,Chapman2009,Furlan2011}. As a result, connecting molecular detections to disk age and C/O evolution has remained challenging. The unprecedented spatial and spectral resolution of JWST-MIRI now enables separation of disk and extended components, even in highly inclined and extinguished systems, offering a way to disentangle disk chemistry from environmental effects.

Against this backdrop, we present new JWST-MIRI observations of ISO-Oph 37 (also  called LFAM 3 or 2MASS J16262357-2424394), a K7 flat-spectrum young stellar object in the $\rho$~Ophiuchi star-forming region. \autoref{tab:source_info} lists a compilation of its observed properties. Previous studies, such as those from the ODISEA ALMA survey \citep{cieza2019}, classify ISO-Oph 37 as a young, disk-bearing YSO with a disk inclination of $\sim$72°, consistent with its flat-spectrum SED. Earlier near-IR surveys by \citet{greene1994} also identify ISO-Oph 37 as a flat-spectrum source, though later classifications \citep{bontemps2001} mark it borderline Class~II, underscoring its ambiguous evolutionary status. More recent and higher sensitivity ALMA observations of its continuum confirm its high inclination (72.6$^\circ$), and found evidence to classify it as a late Class~I in evolutionary terms \citep{vioque2025}. Its flat SED, high disk inclination, and classification uncertainties imply that ISO-Oph 37 is still relatively young ($\lesssim$1\,Myr old) and possibly retains some envelope material \citep{ruiz-rodriguez2025}, likely transitioning between the embedded protostellar and more evolved T Tauri stages.

The JWST/MIRI spectrum of ISO-Oph~37, a solar-type source, reveals strong absorption by \ce{C2H2}, \ce{HCN}, \ce{CO2}, \ce{CH4}, and \ce{H2O}, while notably lacking detectable absorption from common ice species. We further complement the MIRI spectrum with high-resolution \textit{Keck}/NIRSPEC observations of CO, providing additional kinematic and thermal constraints through the detection of both disk emission and an absorption component. Although its flat-spectrum classification and possible residual envelope suggest that the source is still young, the absence of ice absorption and the prominence of acetylene and methane more closely resemble the chemistry of more evolved, carbon-rich disks. This apparent mismatch between evolutionary classification and chemical inventory makes ISO-Oph~37 a valuable test case: it highlights the importance of geometry and local environment in shaping molecular spectra, and it suggests that the relationship between hydrocarbon abundance and disk age in solar-mass stars may be more complex than a simple evolutionary sequence. This paper is organized as follows. In \autoref{sec:observations} we describe the JWST/MIRI and \textit{Keck}/NIRSPEC data acquisition and reduction. In \autoref{sec:methods} we outline the modeling of both data sets, and in \autoref{sec:results} we present the results. In \autoref{sec:discussion} we examine the origin of the absorption spectra, the carbon budget of the system, and the broader context of carbon evolution in planet-forming regions. Finally, our conclusions are summarized in \autoref{sec:conclusion}.

\begin{figure*}[ht!]
\includegraphics[width=18cm]{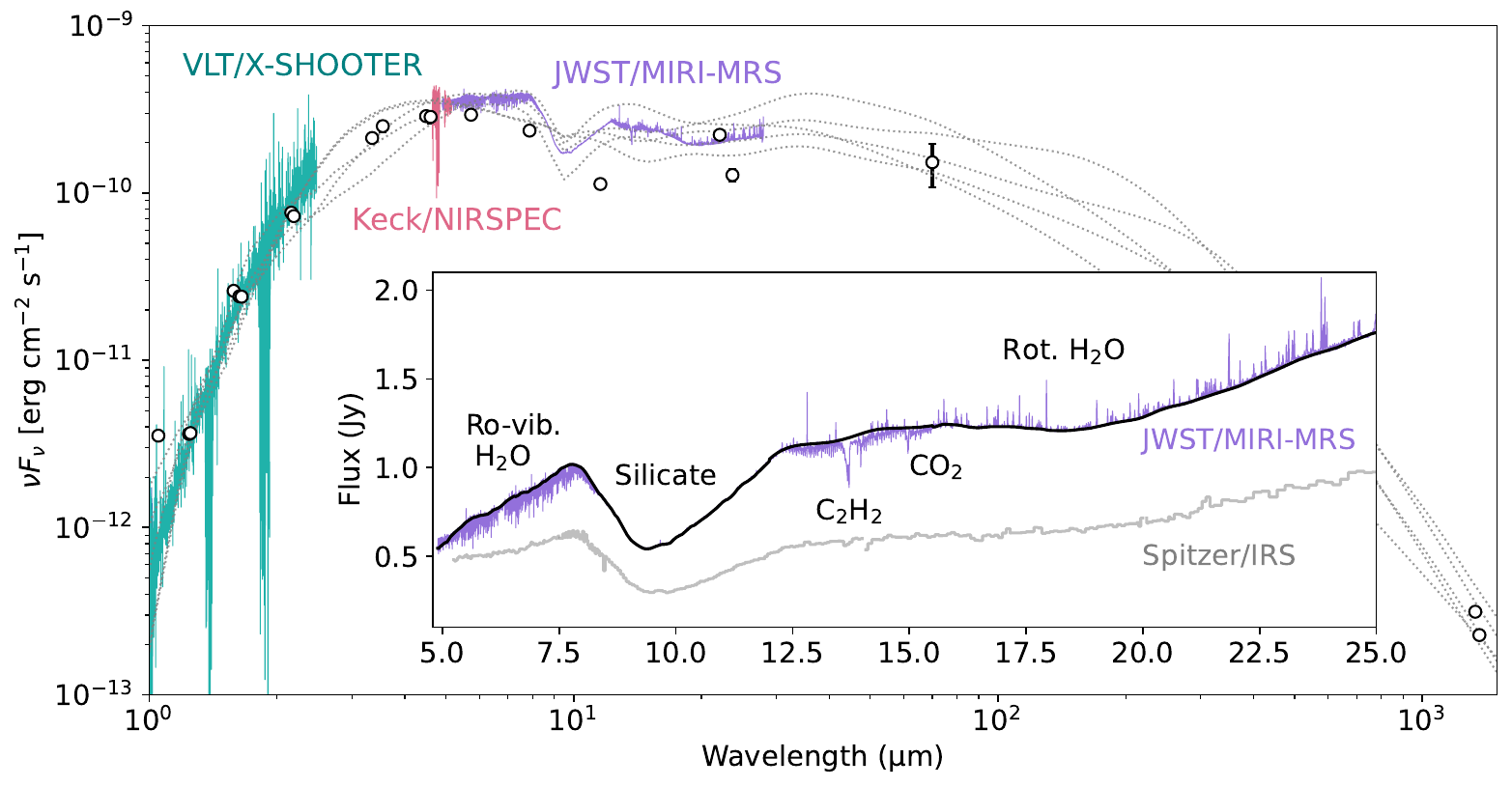}
\caption{Spectral energy distribution of ISO-Oph 37. Photometric points are compiled from \citet{zhang2025,marton2024}. The X-SHOOTER spectrum is taken from \citet{manara2015}. The dotted gray lines show the best-fit SEDs, described in \autoref{app:sed_modeling}. The inset shows the JWST/MIRI-MRS spectrum, and \textit{Spitzer}/IRS spectrum taken from the CASSIS database \citep{CASSIS_lowres}, with some identified gas-phase features. The black line shows the inferred continuum. The shown SED is not corrected for extinction.}
\label{fig:full_spec}
\end{figure*}

\begin{deluxetable}{ccc}[ht!]
\tablecaption{Source properties}

\tablehead{\colhead{Property} & Value & \colhead{Reference} }
\startdata
Spectral Type & K7 & \citet{manara2015} \\
$L_{\rm bol}$ & 1.82\,$L_\odot$ &\citet{Evans09}\\ 
$T_{\rm bol}$ & 670\,K&\citet{Evans09}\\ 
$\alpha_{\rm IR}$ & -0.01& \citet{Evans09}\\
$M_{\star}$ & 0.6\,$M_\odot$& \citet{ruiz-rodriguez2025}\\
$R_{95}$ & 104 au& \citet{vioque2025}\\
Distance & 138.4 pc& \citet{ortiz-leon2018}\\
$A_{\rm v}$ & 16.1 & \citet{manara2015}\\
$i$ & 72.6$^\circ$& \citet{vioque2025}\\ 
\enddata
\end{deluxetable}
\label{tab:source_info}

\section{Observations} \label{sec:observations}

\subsection{JWST/MIRI-MRS}\label{sec:miri_observations}

ISO-Oph 37 was observed with JWST/MIRI-MRS \citep{wright2023,wells2015,argyriou2023} on March 16th 2024 as a part of the GO 3034 program (P.I. K. Zhang) with all three MRS gratings and a total exposure of 455.1\,s per sub-band. The spectrum was reduced with the JDISCS pipeline (v9.1; \citealt{pontoppidan2024}) to maximize signal-to-noise, reduce fringing, and improve the wavelength calibration. The details of the pipeline can be found in the original publication. Briefly, relative to the standard MRS pipeline (run through Stage~2b), we use the JDISCS pipeline to replace the default fringe/throughput calibration with an empirical spectral response function derived from an asteroid calibrator, yielding improved fringe removal and S/N. For v9.1, the 1.20.2 version of the JWST Calibration software is used \citep{bushouse2025}, along with the jwst\_1464.pmap Calibration Reference Data System context. The 1D spectrum is extracted with a wavelength-dependent aperture ($1.4\times1.22\,\lambda/D$) for each sub-band. The final extracted spectrum can be seen in \autoref{fig:full_spec}. Line images are generated by following the approach described in M. Narang et al. (2026, submitted). In short, the central point source is subtracted by a point-spread function (PSF) measured in 6 velocity channels adjacent to the line. To account for spatial drift in the PSF centroid as a function of wavelength, only PSF channels with similar centroid to the line are selected. The combined PSF is scaled to the peak of the flux of the central source in each wavelength channel before subtraction. This efficiently removes the wings of the PSF from extended structure. The 0th moment line map is calculated by integrating over the full lines to $<$90\% of the peak flux density. 

The spectrum of ISO-Oph 37 exhibits a mix of emission and absorption features, with a particularly dense clustering of ro-vibrational water absorption lines between 5--8\,\micron. However, unlike typical protostars or more evolved disks seen edge-on \citep{sturm2023_ices,rocha2025}, ISO-Oph 37 lacks strong \ce{H2O} or \ce{CO2} ice absorption features. Instead, it shows only silicate absorption at 10\micron, suggesting that the line of sight predominantly samples warm material rather than cold envelope or disk midplane gas. The 12-16\,\micron~ region is also dominated by strong C$_2$H$_2$ absorption. At high column densities, emission from this molecule can form a pseudo-continuum, complicating the separation of gas and dust contributions \citep{tabone2023, kanwar2024,kaeufer2024}. To avoid overestimating the C$_2$H$_2$ column density by assuming a specific contribution from the continuum, we implemented a two-part strategy for estimating the continuum, tailored to different wavelength regimes.

Outside the 12-16\,\micron\,region, we implemented the continuum-subtraction algorithm from Zhang et al. (in prep), adapted from the empirical methods of \citet{pontoppidan2024} and \citet{banzatti2024}. The continuum is determined by iteratively smoothing the spectrum over fixed wavelength intervals to trace the underlying dust emission and interpolating between continuum-dominated points to construct a smooth baseline. This approach removes broad dust features while preserving narrow gas lines, though in regions of dense line blending it may subtract a small fraction of the gas signal. To mitigate this effect, we excluded specific intervals from the fit, namely, 7.55-7.75\,\micron, where CH$_4$ absorption is expected, and used different smoothing window sizes and numbers of iterations (i.e., the number of times the smoothing and interpolation are repeated) at short and long wavelengths.

For the 12-16\,\micron\, region, where the presence of C$_2$H$_2$ emission can significantly alter the apparent continuum, we simultaneously modeled the continuum and gas absorption using a third-degree spline function. This approach, detailed further in Section~\ref{sec:slab_models}, ensures a more accurate recovery of the underlying dust continuum. The final continuum model across the full spectral range is shown in the inset of \autoref{fig:full_spec} as the solid black line.

\begin{figure*}[ht!]

\includegraphics[width=18cm]{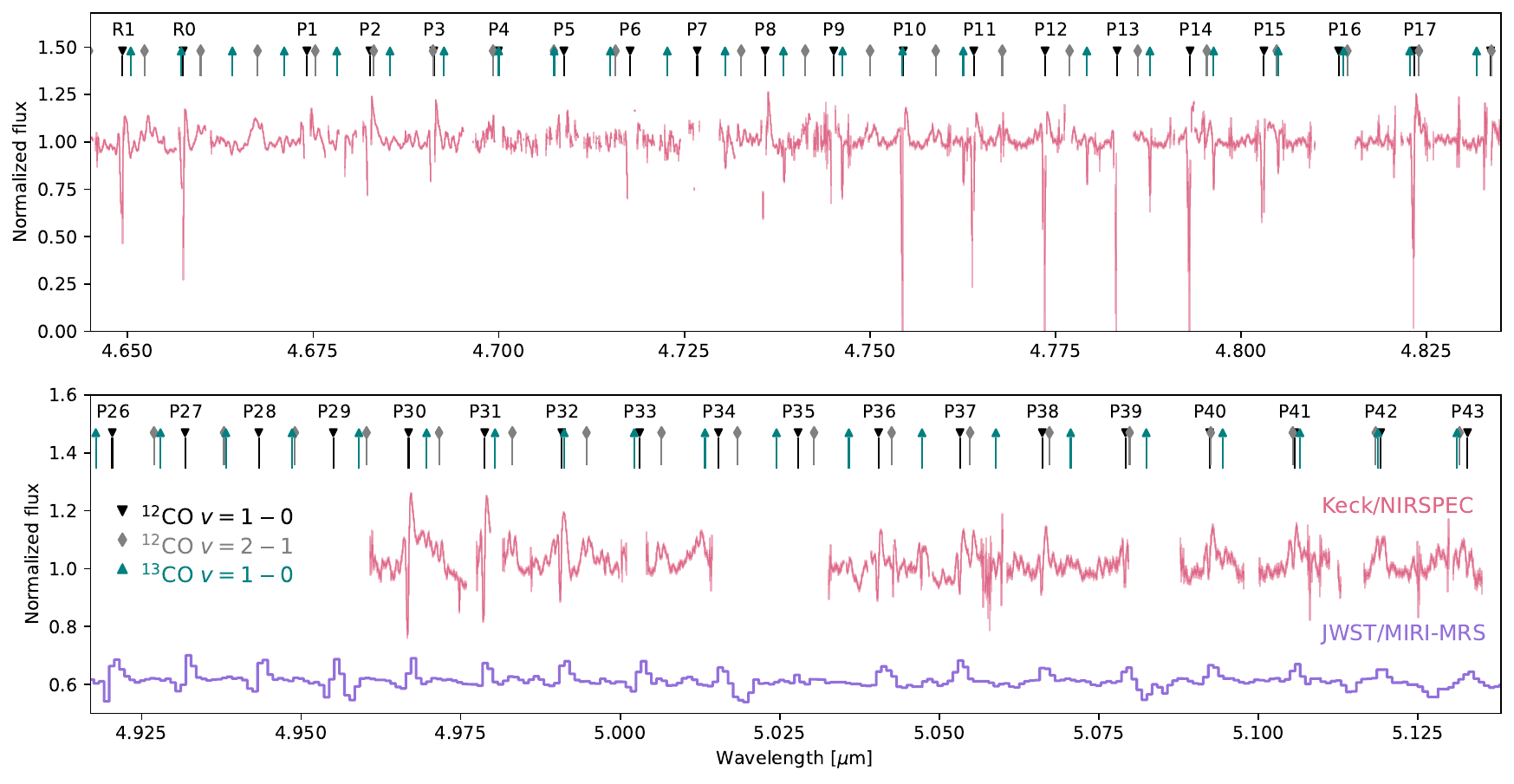}
\caption{Observed CO lines in ISO-Oph 37. The pink line shows the \textit{Keck}/NIRSPEC spectrum and the purple line shows the JWST/MIRI-MRS data. The MRS spectrum is shifted down for clarity. Vertical lines and makers show the locations of the \ce{^12CO} $v=1-0$ and $v=2-1$, and \ce{^13CO} $v=1-0$ lines. Labels denote individual \ce{^12CO} $v=1-0$ rotational quantum numbers.}
\label{fig:co_nirspec}
\end{figure*}

\subsection{\textit{Keck}/NIRSPEC}

To account for the full reservoir of carbon and oxygen, we also include \textit{Keck}/NIRSPEC observations of the $4.6-5.2$~\micron\,region covering multiple $^{12}$CO and $^{13}$CO transitions. High dispersion M-band spectra were acquired over several epochs from 2023-2025 with the updated NIRSPEC2.0 cross-dispersed echelle spectrograph \citep{McLean1998, Martin2018} on the Keck II Telescope in its adaptive optics mode, using R-band light from the central star. The total integration time on source was 144 minutes, over programs on 2023 June 24, 26, 27 (PID: C211; PI: G. Blake), 2023 July 11 (PID: H226; PI: J. Williams), 2024 April 18 (PID: C421; PI: G. Blake), and 2025 June 13, 14 (PID: H397; PI: A. Anderson). Conditions were generally clear, with K-band seeing ranging from 0.$''$4 to 1.$''$2 and an average precipitable water vapor (PWC) of 2 mm (and a total PWV range of 0.6-6 mm). 

On all 2023-2024 nights the 0.$''$027 $\times$ 2.$''$26 slit was used, yielding a resolving power of R$\sim$37,500 ($\sim$8~km\,s$^{-1}$), while those in 2025 used the 0.$''$041 $\times$ 2.$''$26 slit, which provided a resolving power of R$\sim$25,000 ($\sim$12 km s$^{-1}$). Because only a modest fraction of the full M band window  can be covered in a single echelle setting with NIRSPEC2.0, two echelle and cross-disperser settings to capture a wide range of CO lines with varying excitation energy, over a wavelength range of 4.64 to 5.16 \micron, but with a gap from 4.85 to 4.95 \micron.

We used an ABBA nodding pattern to facilitate the removal of sky and background fluxes we employed an ABBA nodding pattern, with each nod having a one minute exposure time. To enable searches for spectro-astrometric signatures in the disk CO gas, the spectrograph’s slit was oriented parallel and anti-parallel to the mm-disk’s major axis at 48/228 degrees and along the disk minor axis at 138/218 degrees \citep{cieza2021}. The bright star HR 5984 ($\beta$ Sco, V4$\sim$2.62 mag; spectral type B1V), served as the primary telluric standard, with maximum separations in airmass between the target and standard of 0.15, with a median separation of $\sim$0.06. HR 5812 was chosen as the telluric standard for the 2024 April 18 data.

\begin{deluxetable*}{lcccccccc}[ht!]
\tablecaption{Parameters from the slab model fitting}

\tablehead{\colhead{Molecule} & Line Type& \colhead{log$_{10}$ $N_{\rm col}$ [cm$^{-2}$]} & \colhead{$T_{\rm ex}$ [K]} & \colhead{$f_c$} & \colhead{log$_{10}$$A_{\rm slab}$ [au$^2$]}&\colhead{Fitting range [\micron]}&}
\startdata
OH - warm & Emission & 16.57$^{+0.03}_{-0.03}$ & 499.38$^{+6.15}_{-6.99}$ &  & 1.34$^{+0.02}_{-0.02}$ & 15.50--27.00 \\
OH - hot & Emission & 16.60$^{+0.19}_{-0.42}$ & 1582.79$^{+298.00}_{-144.60}$ &  & -0.74$^{+0.06}_{-0.05}$ & 15.50--27.00 \\
\ce{H2O} - cold & Emission & 16.97$^{+0.02}_{-0.02}$ & 208.11$^{+1.07}_{-0.99}$ &  & 2.50$^{+0.00}_{-0.01}$ & 18.8--27.00 \\
\ce{H2O} - warm & Emission & 18.22$^{+0.01}_{-0.01}$ & 437.98$^{+3.20}_{-0.25}$ &  & 0.74$^{+0.01}_{-0.01}$ & 15.5--27.00 \\
\ce{H2O} & Absorption&19.11$^{+0.02}_{-0.02}$ & 605.28$^{+5.81}_{-5.19}$ & 1.00$^{+0.00}_{-0.00}$& &4.91--9.00\\ 
\ce{CO2} & Absorption&16.99$^{+0.01}_{-0.02}$ & 420.39$^{+6.56}_{-3.96}$& 1.00$^{+0.00}_{-0.01}$& &12.00--16.00\\ 
\ce{C2H2} & Absorption&17.61$^{+0.03}_{-0.02}$ & 620.67$^{+6.44}_{-2.70}$& 0.57$^{+0.01}_{-0.02}$& &12.00--16.00\\ 
\ce{^{13}CCH2} & Absorption&16.78$^{+0.14}_{-0.08}$ & 681.97$^{+6.86}_{-5.64}$& 0.47$^{+0.02}_{-0.02}$&&12.00--16.00\\
\ce{HCN} & Absorption&16.60$^{+0.03}_{-0.03}$ & 461.79$^{+10.79}_{-12.26}$& 1.00$^{+0.00}_{-0.01}$& & 12.00--16.00\\
\ce{CH4} & Absorption&18.42$^{+0.04}_{-0.10}$& 550.25$^{+46.96}_{-14.32}$&  0.59$^{+0.03}_{-0.02}$& & 4.91--16.00\\
CO & Absorption &18.34$^{+0.56}_{-1.18}$&  962.19$^{+568.75}_{-503.15}$& [1.00] & &4.96--5.14\\
\enddata
\tablecomments{The brackets indicate parameters that were fixed during the fitting.} 
\end{deluxetable*}
\label{tab:slab_vals}

As described in \citet{Anderson2024A}, the data were reduced and calibrated using a processing pipeline that performs standard infrared data reduction techniques (i.e., flat-fielding, dark-subtraction, bad-pixel masking, A-B pair subtraction). Specifically, sequential A-B pairs from a given night that were oriented along the same slit position angle were stacked prior to wavelength calibration, standard star division, and 1D extraction to increase the signal-to-noise ratio of the target. These 1D spectra were extracted by fitting a polynomial to each trace in the 2D spectrum. Next, the Reference Forward Model, a line-by-line radiative transfer model designed to simulate telluric infrared spectra \citep{Dudhia2017}, was used to calibrate the wavelength/plate scale. The stacked A-B pairs so processed were then divided by the standard star’s spectrum to remove telluric absorption features. Regions of the spectrum with a sky transmission value of less than 0.7 (as indicated by a normalized standard-star spectrum) were not included in our analysis to reduce the possibility of interpreting a telluric absorption feature as a spectral feature from the star/disk.

After wavelength calibration, the spectra were shifted to the stellar reference frame using an ISO-Oph37 radial velocity of -7.9 km s$^{-1}$ \citep{sullivan2019} to perform the heliocentric velocity correction needed to bring all epochs into a common reference frame (including the Earth-induced shifts at the time of the observations). The full suite of data were stacked and weighted according to errors, and a spline fit was used to generate a continuum-subtracted spectrum. Finally, the total flux beyond 5 \micron~ was fixed to that derived in the JWST-MIRI spectrum. The calibrated CO spectrum, along with the MIRI observations, are shown in \autoref{fig:co_nirspec}. Multiple transitions of \ce{^12CO} across two vibrational transitions, and multiple rotational transitions, are identified in absorption and emission, along with lines of \ce{^13CO} predominantly in absorption. We also detect a couple of rovibrational water lines, as discussed in \autoref{sec:co_analysis}.

\section{Methods} \label{sec:methods}

\subsection{MIRI slab modeling} \label{sec:slab_models}
We derived the physical parameters of the observed emission and absorption lines using local thermodynamic equilibrium (LTE) slab models. These models calculate the total optical depth of a line from an assumed excitation temperature ($T_{\rm ex}$) and column density ($N_{\rm col}$). For emission features, the observed flux can also be determined by assuming an emitting area ($A_{\rm em}$) and scaling the output to the source distance. LTE slab modeling has been extensively applied in the literature to extract physical parameters from both emission and absorption features \citep{arulanantham2025,temmink2025,vangelder2024}, and it has been shown that the retrieved parameters are usually in agreement with more complex thermochemical models \citep{vlasblom2025}. Our observed spectrum contains a combination of emission and absorption components, requiring a region-by-region fitting strategy. Hence, we first separated the features into emission and absorption sets, and modeled them independently. Throughout this work, we generate the slab model spectra using the \texttt{spectools\_ir} package \citep{spectools_ref}, which uses molecular data from the HITRAN database \citep{gordon2026_hitran}.

\subsubsection{Emission}

For the emission component, present beyond 15.5~\micron, we adopt the same fitting algorithm as described in Zhang et al. (in prep). Briefly, the slab models are computed in two MIRI/MRS wavelength intervals (15.5-18.8 and 18.8-27\,\micron), convolved to the corresponding MRS spectral resolution \citep{pontoppidan2024,banzatti2024}, and compared to the continuum-subtracted observed spectrum. We adopt a Gaussian likelihood, evaluated as the sum over the two spectral intervals,
\begin{equation}
\ln \mathcal{L}(\boldsymbol{\theta}) = -\frac{1}{2}\sum_{k=1}^{2}\sum_{\lambda_i \in [\lambda_{k,\min},\lambda_{k,\max}]}
\left[\frac{F_{\rm obs}(\lambda_i)-F_{\rm mod}(\lambda_i|\boldsymbol{\theta})}{\sigma_{\rm obs}(\lambda_i)}\right]^2,
\end{equation}
where $F_{\rm obs}(\lambda)$ is the continuum-subtracted spectrum (shifted by $v_r=-7.9$\,km\,s$^{-1}$, \citet{sullivan2019}) and $F_{\rm mod}(\lambda|\theta)$ is the summed molecular slab model in the same interval, evaluated at the observed wavelengths. The set of molecules included in $F_{\rm mod}$ is allowed to differ between intervals, and \autoref{tab:slab_vals} lists what wavelength interval was used for each species. 
Here $\sigma_{\rm obs}$  represents the flux uncertainty of the observed spectrum, for which we adopt uncertainties of 2.5, 5, and 8\,mJy for the 15.5--20, 20--22, and 22--27\,\micron\ regions, respectively, as calculated from the flux standard deviation of line-free regions. The intrinsic line broadening of the slab models is assumed to arise from thermal motions and therefore varies with the temperature of each species. Rather than correcting the observation for extinction, we extinct the models before comparing them with the observed spectrum. Similar to other sources with absorption \citep{vangelder2024a}, we use the absolute extinction law from \citet{McClure2009}, assuming $A\rm_V=16$ \citep{manara2015}. The posterior is sampled with an affine-invariant Markov Chain Monte Carlo (MCMC) ensemble sampler. We adopt uniform priors within the parameter bounds listed in \autoref{tab:priors}. For the cold water component (\ce{H2O}$_{\rm cold}$), we additionally impose a Gaussian prior on temperature centered at $T_{\rm ex}=200$\,K with $\sigma=50$\,K, while all other parameters retain uniform priors.

We initially adopted three H$_2$O slabs representing different temperature reservoirs ($\sim$200\,K, 400\,K, 900\,K) that have been identified in the analysis of T Tauri stars \citep{banzatti2024}. However, the hot water component contributed negligibly to the model flux and is therefore excluded from the final fit. We also detected highly excited OH emission between 9-11\,\micron, likely produced by H$_2$O photodissociation. This component, typically not in LTE, can also contribute to the OH emission observed at longer wavelengths. To account for these potential contributions, we used the OH models from \citet{tabone2024} as a template and subtracted it from the 8.8-18\,\micron\,spectrum before performing the fitting. For the wavelengths longward of 15\,\micron, we expect the remaining OH emission to be in LTE. 
However, a single LTE OH component could not simultaneously reproduce the remaining features; therefore, we used two OH components, as in previous studies \citep[e.g.][]{schwarz2024}.

The H$_2$O and OH emission lines were simultaneosuly fit between 15.5--27\,\micron, with the cold H$_2$O component included only longward of 18.8\,\micron, where its contribution is expected to be strongest. We do not attempt to fit the CO lines detected with MIRI, as the profiles are a complex combination of emission and absorption. We instead perform the CO analysis on the NIRSPEC observation, detailed in \autoref{sec:co_analysis}. We run the MCMC fitting routine using 100 walkers and $10^4$ steps to ensure convergence. We report the parameter values in \autoref{tab:slab_vals}, corresponding to the marginalized posterior medians (50th percentiles). Quoted uncertainties are the central 68\% credible intervals (16th-84th percentiles).

\subsubsection{Absorption}

\begin{figure}[ht!]
\includegraphics[width=0.48\textwidth]{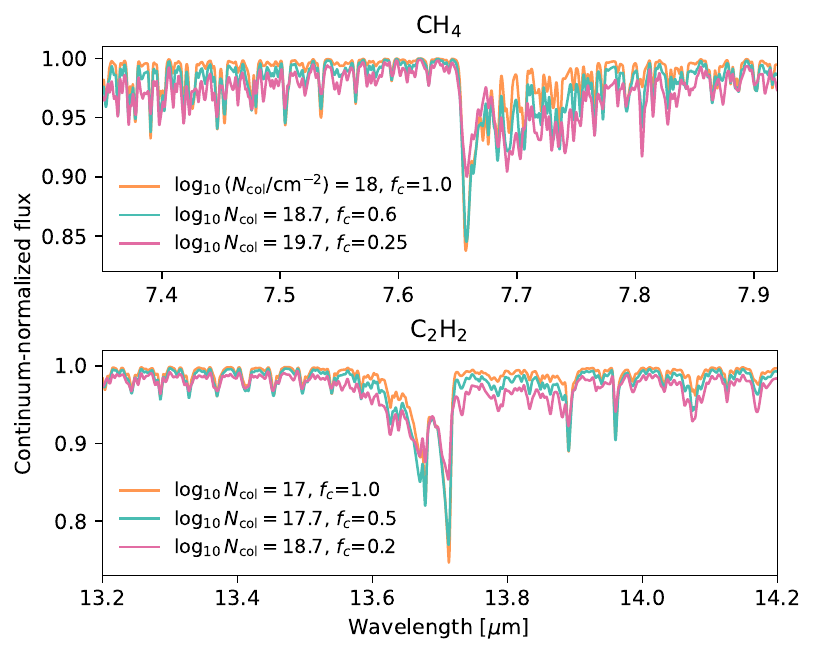}
\caption{Comparison of the effect of the covering fraction $f_c$ on the slab model spectra for CH$_4$ and C$_2$H$_2$. As the column density increases, the flux of the $P$ branch also increases, while the $Q$ branch flux decreases. All models assume $T_{\rm ex}=500$\,K.} 
\label{fig:fc_comparison}
\end{figure}

For the  molecular absorption, present between 4.9--9~\micron~ and 12--15.5~\micron, we implemented two different fitting methods. Unlike in emission, there is no physically meaningful emitting area for absorption. Because the spatial distribution of the absorbing gas relative to the continuum source is generally unknown, we allow the absorber to cover only a fraction of the continuum. We account for this geometry by introducing a covering fraction $f_c$ \citep{jli2024}, that is able to scale down the flux. The absolute flux can then be written as

\begin{equation}
    F_\nu = F_c(1-f_c(1-e^{-\tau_\nu}))\,,
\end{equation}

\noindent where $F_\nu$ is the absolute flux, $F_c$ is the continuum flux, and $\tau_\nu$ is the optical depth of the line. In the absorption slab model, $\tau_\nu$ is computed self-consistently from the excitation temperature and column density of the gas, while the covering fraction $f_c$ is treated as an additional free parameter. We note that while the covering fraction and column density can appear partially degenerate in the optically thin limit, this degeneracy is mitigated when multiple transitions spanning a range of excitation energies are considered. Increasing the column density does not simply deepen existing absorption features, but instead alters the relative strengths of transitions in the $P$, $Q$ and $R$ branches. In contrast, varying the covering fraction primarily rescales the depth of all absorption features without changing the relative line ratios. As a result, models with high column density and low $f_c$ cannot reproduce the observed line ensemble in the same way as models with lower column density and higher $f_c$. The simultaneous fitting of multiple transitions therefore provides independent constraints on $N_{\rm col}$ and $f_c$. To illustrate this point, we show in \autoref{fig:fc_comparison} three representative absorption models with comparable line depths but different combinations of column density and covering fraction. Although the overall absorption strength is similar, the models differ in the relative strengths and presence of transitions across the $P$, $Q$ and $R$ branches, demonstrating that the degeneracy between $N_{\rm col}$ and $f_c$ is not global when multiple lines are considered.

The model spectrum that is compared to the observations is the absolute flux $F_{\nu}$, constructed by multiplying the continuum-normalized absorption model by the continuum flux $F_c$. As a result, uncertainties in the continuum determination propagate directly into the modeled absorption depths, making an accurate estimate of $F_c$ essential for robust constraints on the parameters. Our continuum treatment is therefore chosen based on whether the local continuum can be reliably defined in the presence of broad absorption features, or whether such features prevent \textit{a priori} continuum placement. In the shorter wavelength region (5.5-9\,\micron), most of the rovibrational water and methane transitions occur at wavelengths shorter than 8\,\micron, where the spectrum is largely unaffected by the deep silicate absorption feature extending from $\sim$8-12\,\micron. We therefore adopt the pre-computed continuum derived from our continuum subtraction algorithm described in \autoref{sec:miri_observations} up to 12\,\micron.

\begin{figure*}[ht!]
\centering
\includegraphics[width=15cm]{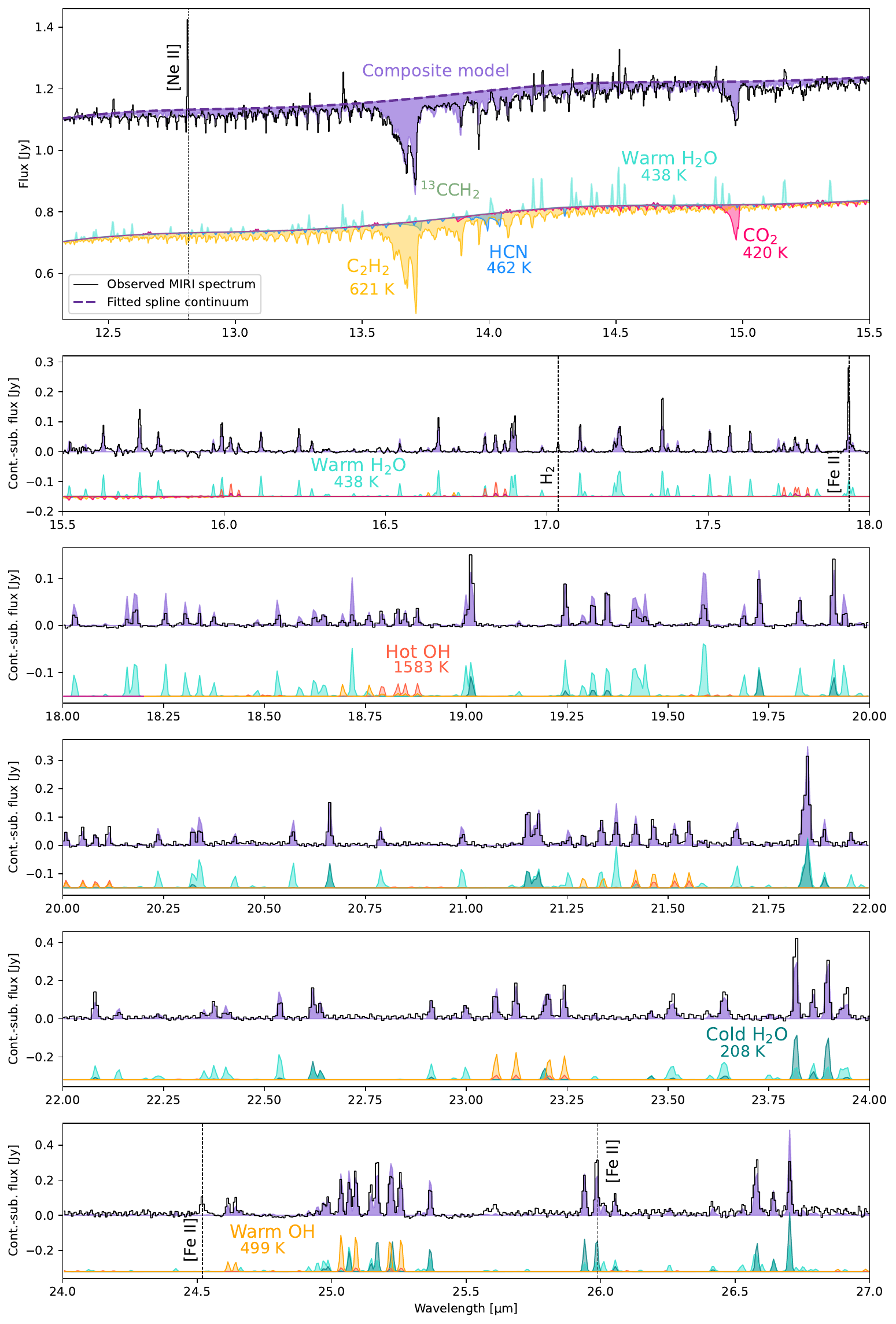}
\caption{Best-fit LTE slab models beyond 12\,\micron. The top panel shows the absolute flux and the fitted continuum, whereas the rest of the panels show the continuum-subtracted flux. Molecular hydrogen and atomic lines are indicated with a vertical dashed line. The contribution from warm H$_2$O in the top panel is shown only for reference, as it was not used in the fitting procedure of the organics. }
\label{fig:lte_fits}
\end{figure*}

\begin{figure*}[ht!]
\centering
\includegraphics[width=15cm]{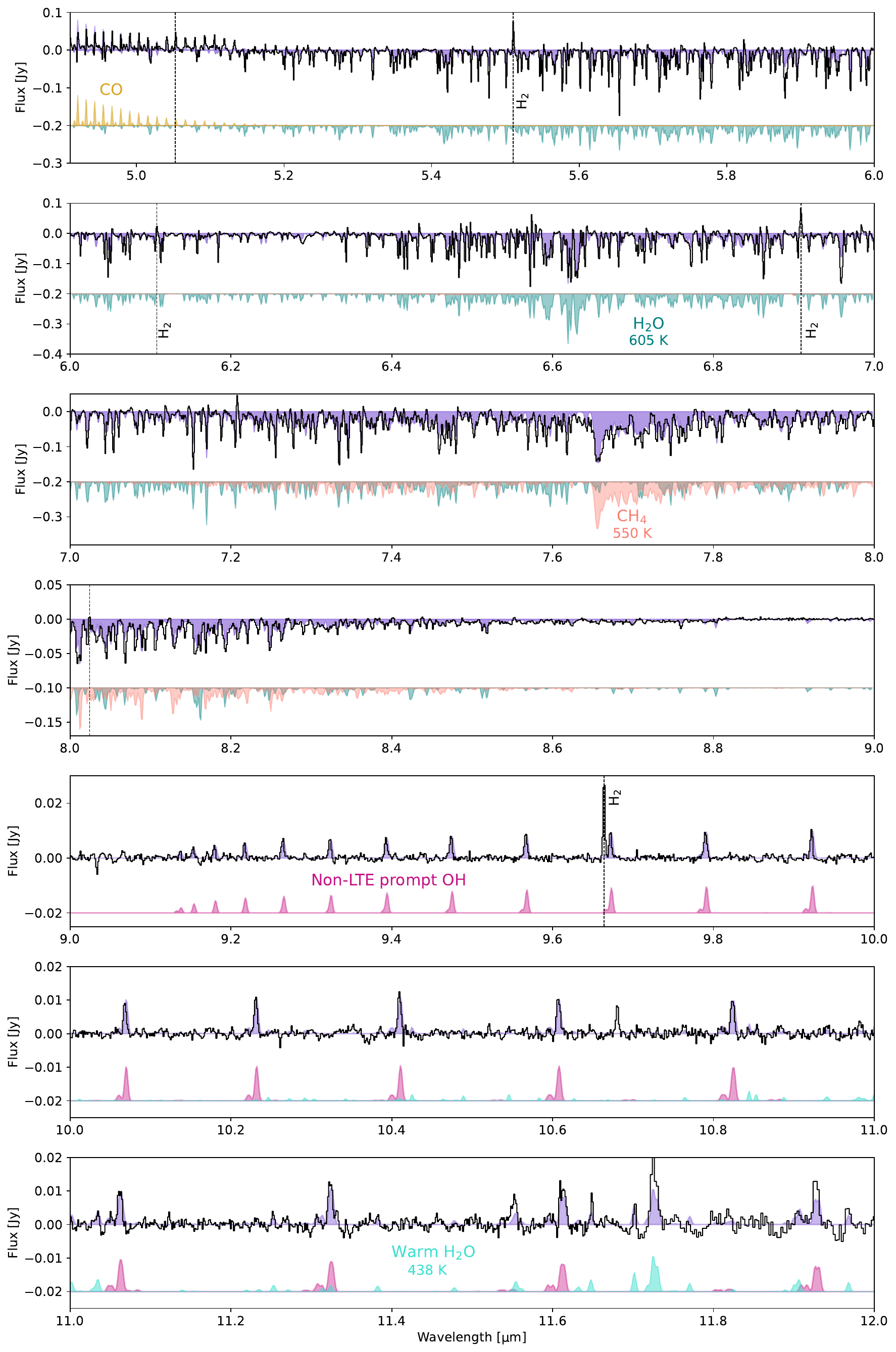}
\caption{Best-fit LTE slab models between 4.91--12\,\micron. Molecular hydrogen and atomic lines are indicated with a vertical dashed line. The non-LTE prompt OH models are taken from \citet{tabone2024}. The warm H$_2$O models were reproduced from the best-fit parameters obtained for longer wavelengths, shown here just for reference. A CO model is shown just for reference.}
\label{fig:short_lte_fits}
\end{figure*}

In the case of the organics in the 12--16\,\micron\, range, the \ce{C2H2} absorption is abruptly ``cut off" by the silicate absorption feature. In addition, because \ce{C2H2} is the strongest feature detected in our spectrum, we consider the possibility of a pseudo-continuum. When the column density of hydrocarbons is too high, these could saturate and create an optically-thick pseudo-continuum that lies below an optically-thin gas component, as has been observed in multiple hydrocarbon-rich sources \citep{tabone2023,kanwar2024,kaeufer2024,arabhavi2024,jang2025}. This feature makes it difficult to correctly estimate a continuum. Therefore, instead of trying to establish a continuum before fitting the gas features, we remain agnostic to the continuum levels and simultaneously fit the gas features together with a spline function. We select eight equally distributed wavelength points between 12--16\micron, and estimate the continuum flux levels at each point by fitting a third-degree univariate spline function. Since the wavelength interval considered is relatively narrow, a third-degree spline provides sufficient flexibility to capture smooth continuum curvature while avoiding the introduction of artificial structure in the residuals. We also note that using a fourth degree spline results in the same retrieved parameters. The eight selected points will create a continuum $F_c$, that is multiplied by each model before the likelihood function is evaluated. Therefore, the continuum is fit simultaneously with the \ce{C2H2}, HCN and \ce{CO2} components, for which we vary the column density, temperature, and covering fraction. We also include a \ce{^{13}CCH2} component, as it is commonly observed in systems with strong \ce{C2H2} \citep[e.g.][]{tabone2023,arabhavi2025}. However, we include a condition in our prior that enforces the covering fraction and temperature to be within 10\% of the values for \ce{C2H2}, under the assumption that both species are roughly co-spatial (see \autoref{app:priors} for more details). The continuum and molecular absorption are therefore fit simultaneously, allowing the fit to accommodate a pseudo-continuum if present. The adopted parameter priors, spline node locations, and best-fit spline coefficients for this region are reported in \autoref{app:priors}.

Similar to the algorithm implemented for emission, the fitting routines for absorption implement an MCMC algorithm to efficiently explore the parameter space, using 100 walkers and $10^4$ steps. Uniform priors and physically motivated constraints are applied to all absorption model parameters, including temperature, column density, covering fraction, and continuum spline coefficients; the full prior specification is described in \autoref{app:priors}. The resulting parameters and their uncertainties are obtained from the posterior distributions, and they are listed in \autoref{tab:slab_vals}. As in the emission analysis, we report posterior medians and 68\% credible intervals. The resulting slab models are shown with the continuum subtracted spectrum in \autoref{fig:lte_fits} and \ref{fig:short_lte_fits}. We note that we only show the \ce{H2O}$_{\rm warm}$ model in the 12--16\,\micron\, region for reference, as it was not considered during the fitting of the absorption features.

\subsection{Keck/NIRSPEC high-resolution CO analysis} \label{sec:co_analysis}

A significant fraction of the carbon and oxygen in disks can be locked in CO, so constraining its abundance is essential for characterizing the overall chemistry of the source. Because the spectral resolution of MIRI-MRS is insufficient to resolve the typically complex, multi-component CO line profiles seen in disks \citep{brown2013,temmink2024}, and the MIRI wavelength coverage does not capture the full CO fundamental but rather just its tail, we complement the JWST data with high-resolution \textit{Keck}/NIRSPEC observations of CO. The NIRSPEC profiles shown in \autoref{fig:co_nirspec} display a combination of emission and blue-shifted absorption. Since the absorption lines are visibly shifted by different velocities with respect to the rest velocity of the line transitions, a standard slab model fitting routine would struggle to minimize the residuals. Instead, we first construct an empirical, pure-absorption spectrum and then fit it with LTE CO slab models.

\begin{figure}[ht!]
\includegraphics[width=8.5cm]{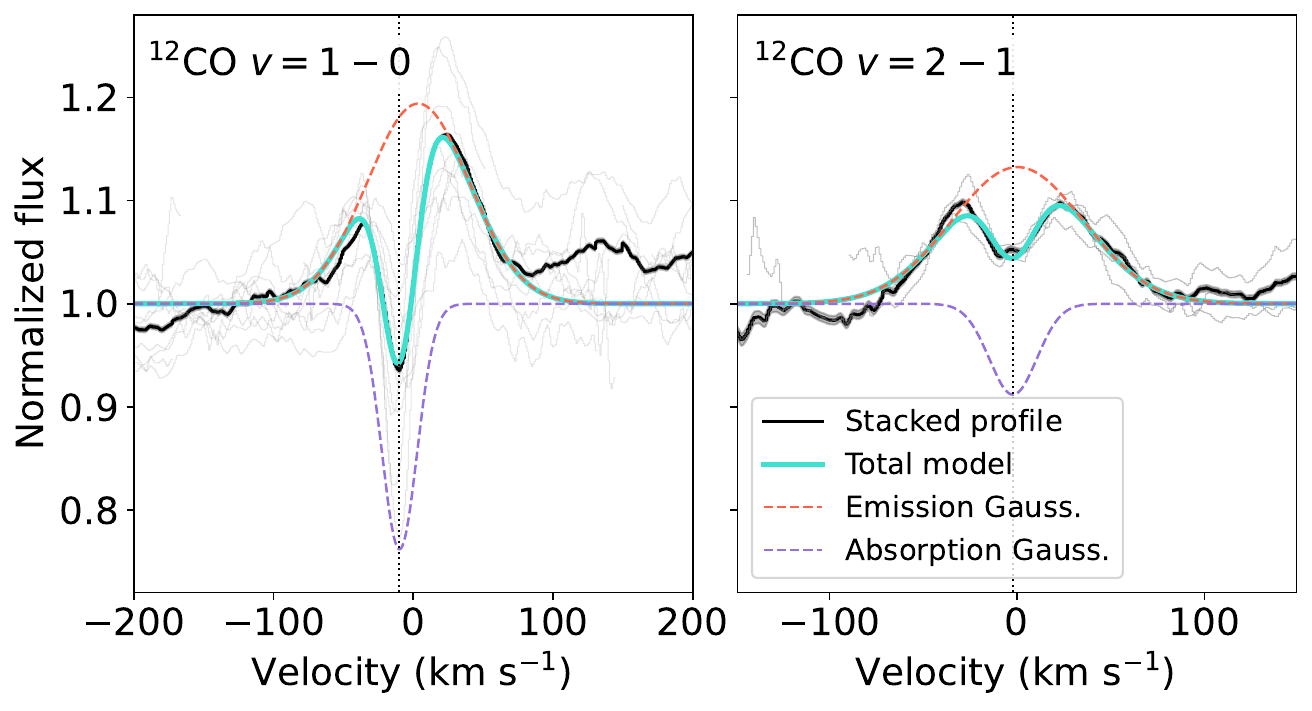}
\caption{Stacked \ce{^12CO} profiles for the NIRSPEC observations. Light gray lines show the individual profiles while the black line shows the final stacked profile. The dotted vertical lines show the velocity centroid of the absorption component, $v_0$ from \autoref{tab:co_gauss_params}.}
\label{fig:co_profiles}
\end{figure}

We modeled the ro-vibrational ${}^{12}$CO absorption by first characterizing the line profile empirically and then fitting it with slab models. Starting from the continuum-normalized spectrum with associated uncertainties, we selected a set of relatively unblended ${}^{12}$CO $v=1$-0 and $v=2$-1 lines from the HITRAN line list  \citep{gordon2026_hitran} in the ranges 4.64--4.84\,\micron~ and 4.95--5.14\,\micron. For each line separately, we shifted the spectrum into the velocity frame of each transition and extracted a velocity segment around the line. We retained only those lines with adequate sampling of the absorption core and without obvious telluric contamination. For the $v=2$-1 sample we additionally rejected lines lying within 10\,km\,s$^{-1}$ of a $v=1$-0 transition, as this latter one is expected to dominate. The retained per-line segments were combined into an inverse-variance-weighted stacked profile (including both the absorption and emission components), yielding an average line profile and corresponding uncertainty. Following the selection procedure, the retained $v=1$-0 transitions included in the stacked profile are P(30)-P(32), P(36)-P(38), and P(40)-P(43), whereas the $v=2$-1 stack consists of the P(26)-P(28) transitions. The stacked profiles, along with their 1$\sigma$ uncertainty, are shown in \autoref{fig:co_profiles}.

Then, the stacked profiles were modeled as the sum of a broad emission component and a narrower absorption component, approximated as
\begin{equation}
    F_{\rm norm}(v) \simeq 1 + G_{\rm em}(v) - G_{\rm abs}(v),
\end{equation}
\noindent where $F_{\rm norm}(v)$ is the continuum-normalized flux, and $G_{\rm em}$ and $G_{\rm abs}$ are Gaussians in velocity space, and each of these Gaussians is described as:

\begin{equation}
G(v;\,A,v_0,\mathrm{FWHM})
= A\,\exp\!\left[-4\ln 2\left(\frac{v-v_0}{\mathrm{FWHM}}\right)^2\right]\,.
\end{equation}

\noindent In this parameterization, $A$ is the Gaussian amplitude, defined relative to the continuum level of unity, such that $A_{\rm em}>0$ produces an emission peak above the continuum and $A_{\rm abs}>0$ produces an absorption dip below the continuum. The parameter $v_0$ is the centroid velocity of the component relative to the systemic velocity of the line, and $\mathrm{FWHM}$ is the full width at half maximum. To avoid bias from the absorption core during the fitting of the Gaussians, we first fit the emission Gaussian only to the line wings, outside a chosen core interval around the line center. This emission Gaussian model was then used as a baseline; subtracting the emission from the stacked profile isolates a ``dip'' in the core, which we fitted with a second Gaussian to determine the absorption amplitude, centroid, and FWHM. This fitting procedure then results in two sets of parameters for the $^{12}$CO $v=1-0$ and $v=2-1$ transitions. The parameter sets include FHWM of the emission and absorption Gaussians, their amplitudes, and their central velocities, listed in \autoref{tab:co_gauss_params}. The fitted Gaussians are shown in \autoref{fig:co_profiles}.

\begin{deluxetable}{lccccc}
\tablecaption{Best-fit Gaussian parameters for the stacked \ce{^{12}CO} line profiles. \label{tab:co_gauss_params}}
\tablehead{
\colhead{Transition} &
\colhead{Component} &
\colhead{$A$} &
\colhead{$v_0$ (km\,s$^{-1}$)} &
\colhead{FWHM (km\,s$^{-1}$)} &
\colhead{$R_{\rm Kep}$ (au)}
}
\startdata
\ce{^{12}CO} $v=1$--0 & Emission   & 0.1940 & $+3.52$  & 86.19 & 0.26 \\
         & Absorption & 0.2381 & $-9.86$  & 29.65 &  \\
\hline
\ce{^{12}CO} $v=2$--1 & Emission   & 0.1325 & $+0.32$  & 82.70 & 0.28 \\
         & Absorption & 0.0881 & $-2.09$  & 29.80 &  \\
\enddata
\tablecomments{The amplitudes $A$ are dimensionless and defined relative to a continuum-normalized flux of unity. The centroid velocities $v_0$ are measured relative to the systemic velocity of the source.}
\end{deluxetable}

As can be seen in the values for the best-fit Gaussians (\autoref{tab:co_gauss_params}), the absorption component of the $v=1$--0 transitions is blueshifted by $\sim$10\,km\,s$^{-1}$ with respect to the system velocity, whereas the absorption in the $v=2$--1 transitions is only blueshifted by $\sim$2\,km\,s$^{-1}$. Given that the best spectral resolution achievable with our \textit{Keck}/NIRSPEC observations is $\sim$8\,km\,s$^{-1}$, the smaller blueshift measured for the $v=2$--1 absorption is therefore only marginally resolved, while the larger blueshift of the $v=1$--0 absorption represents a more robust, spectrally resolved offset. Furthermore, an absorption component not physically associated with the disk would not be expected to appear in higher vibrational transitions, as this would require very high excitation temperatures and column densities. We therefore interpret the $v=2$--1 profiles as being dominated by the double-peaked line shape characteristic of Keplerian rotation in the disk, rather than tracing the same absorbing material seen in the $v=1$--0 transition. Under this assumption, we estimate a characteristic Keplerian radius for the emitting gas from the FWHM of the emission Gaussian fitted to the $v=1$--0 and $v=2$--1 transitions according to

\begin{equation}
R_{\rm Kep}
= G M_\star\bigg(\frac{\sin i}{\rm FWHM/2}\bigg)^2,
\end{equation}
where $M_\ast$ and $i$ are the stellar mass and disk inclination, respectively,
adopted from \autoref{tab:source_info}. The resulting radii are reported in
\autoref{tab:co_gauss_params}.

Because the absorption seen in the $v=2$--1 profile is unlikely to trace the same
high-velocity absorbing material as the $v=1$--0 lines, we restrict our subsequent analysis and column density determinations to the absorption observed in the $v=1$--0 transition. To construct an empirical CO absorption spectrum that combines multiple lines, we first adopt the emission and absorption FWHM values derived from the stacked profiles as global line widths applicable to all transitions.

With the line widths fixed, we then fit each individual line using a two-Gaussian model consisting of an emission and an absorption component, allowing only the amplitudes and centroid velocities of both components to vary. We limit this procedure to the subset of lines that were previously identified as suitable for inclusion in the stacked profiles.

From the set of successful per-line fits, we retain the absorption amplitudes and centroid velocities and reconstruct an empirical absorption-only spectrum over the full wavelength range by summing the Gaussian absorption contributions from all lines. This procedure yields a synthetic absorption flux spectrum, which we convert into an empirical optical-depth spectrum via
\begin{equation}
    \tau_{\rm Gauss}(\lambda) = -\ln F_{\rm abs,norm}(\lambda),
\end{equation}
where $F_{\rm abs,norm}(\lambda)$ is the continuum-normalized absorption flux.

\begin{figure*}[ht!]
\includegraphics[width=18cm]{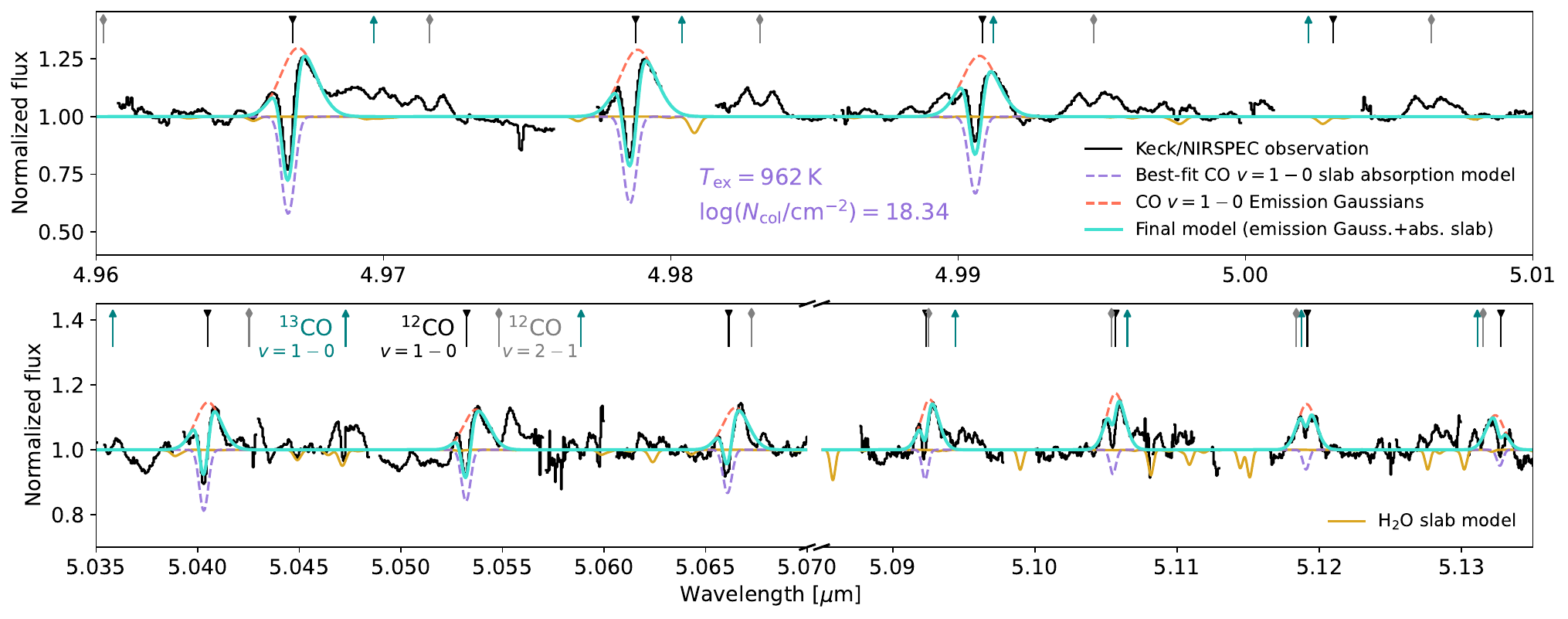}
\caption{Resulting CO profiles from the LTE slab fitting. The purple dashed line shows the best-fit absorption slab model for the CO $v=1$--0 transitions, each line has been velocity shifted to match the absorption centroid. Transition line centers are shown for reference at the top. The $^{13}$CO lines are analyzed in \autoref{app:13co}. Some additional absorption features not associated with CO are consistent with water lines, using a slab model constructed with the same best-fit parameters derived from the MIRI observations (see text).}
\label{fig:co_slabs}
\end{figure*}

Then, to infer the physical state of the high-velocity CO absorption component, we fit our empirical absorption-only spectrum with an LTE slab model. We adopted the observed FWHM of the absorption Gaussian (29.65\,km\,s$^{-1}$, \autoref{tab:co_gauss_params}) as the FHWM for the convolution kernel in our slab models. This choice essentially reflects the fact that the observed FWHM is itself the convolution of the instrumental line-spread function and the intrinsic velocity dispersion of the absorbing gas. We performed an MCMC retrieval on $\tau_{\rm Gauss}(\lambda)$, comparing it to the model optical-depth spectra over the wavelength regions covered by the selected lines. The MCMC explores temperature and column density as free parameters, using the same general slab-model framework as for the rovibrational water and \ce{CH4} analysis described in \autoref{sec:slab_models}, but fixing the covering fraction to $f_c=1$. From the resulting posterior distribution, we adopt as ``best-fit'' parameters the maximum-a-posteriori (MAP) temperature and column density. Uncertainties are estimated from the marginalized posteriors as the 16th-84th percentile range. In this case, we choose to report MAP values rather than posterior medians primarily because the temperature posterior is non-Gaussian (see \autoref{fig:co_posterior}), such that the median can fall in a relatively low-probability region. For the column density, the posterior is more sharply peaked and the MAP and median are very similar, so this choice has a negligible effect. The resulting best-fit parameters are listed in \autoref{tab:slab_vals}, and the slab and overall modeled spectrum are shown in \autoref{fig:co_slabs}. We note that we also detect \ce{^{13}CO} lines in the NIRSPEC spectrum. However, these lines are in pure absorption and trace a much colder ($\sim$30\, K) component, that is likely associated with a more extended envelope/cloud. We present the analysis of the \ce{^{13}CO} lines in \autoref{app:13co}. Finally, some of the absorption features that do not correspond to CO are consistent with water lines. To investigate this, we generate a slab model using the best-fit parameters for the water absorption derived from the MIRI observations (\autoref{tab:slab_vals}) and convolve it to the \textit{Keck}/NIRSPEC resolution. The resulting model is shown in \autoref{fig:co_slabs}, where we find good agreement with several features, particularly near 5.045, 5.0625, and 5.128~\micron. We do not pursue further analysis of these lines, as they are difficult to disentangle from overlapping emission, and the model already provides a satisfactory reproduction of the observed features.

\section{Results}\label{sec:results}

\subsection{Absorption}

Across the combined \textit{Keck}/NIRSPEC and \textit{JWST}/MIRI-MRS wavelength coverage, we detect six molecular species in absorption: CO, \ce{C2H2}, HCN, \ce{CO2}, \ce{H2O}, and \ce{CH4}. Within the MIRI range, these species trace two distinct temperature components: HCN and \ce{CO2} are associated with cooler gas ($\sim$450\,K), while \ce{C2H2}, \ce{H2O}, and \ce{CH4} trace warmer material ($\sim$600\,K). The best-fit column densities (log $N$/cm$^{2}$: \ce{C2H2}=17.61, HCN=16.60, \ce{CO2}=16.99, \ce{CH4}=18.46, \ce{H2O}=19.13) place the source firmly within the “warm inner-disk” regime \citep{arulanantham2025}. The CO absorption traces gas of comparable column density (log $N$(CO)/cm$^{2}$=18.07) but at significantly higher temperature ($\sim$1000\,K). This difference likely reflects the fact that the CO slab fitting relies on higher-energy transitions, as lower-energy lines that would probe cooler gas are strongly affected by telluric contamination. Although column density ratios can be derived from these measurements, the differences in temperature indicate that not all species necessarily arise from the same physical region. In particular, the higher excitation temperature of CO suggests that it may trace material located closer to the star than the species detected with MIRI. Accordingly, in the remainder of this work we present ratios relative to CO for completeness, but focus our interpretation on ratios involving species with comparable excitation temperatures (i.e., the rest of the species). Within this framework, we find that the absorbing gas is enriched in hydrocarbons, with (\ce{C2H2}+\ce{CH4})/CO$\approx2$ and (\ce{C2H2}+\ce{CH4})/HCN$\approx80$. In terms of the oxygen budget, most oxygen is stored in water, whose column density exceeds that of CO by an order of magnitude and that of \ce{CO2} by nearly two orders of magnitude.

\begin{figure*}[ht!]

\includegraphics[width=18cm]{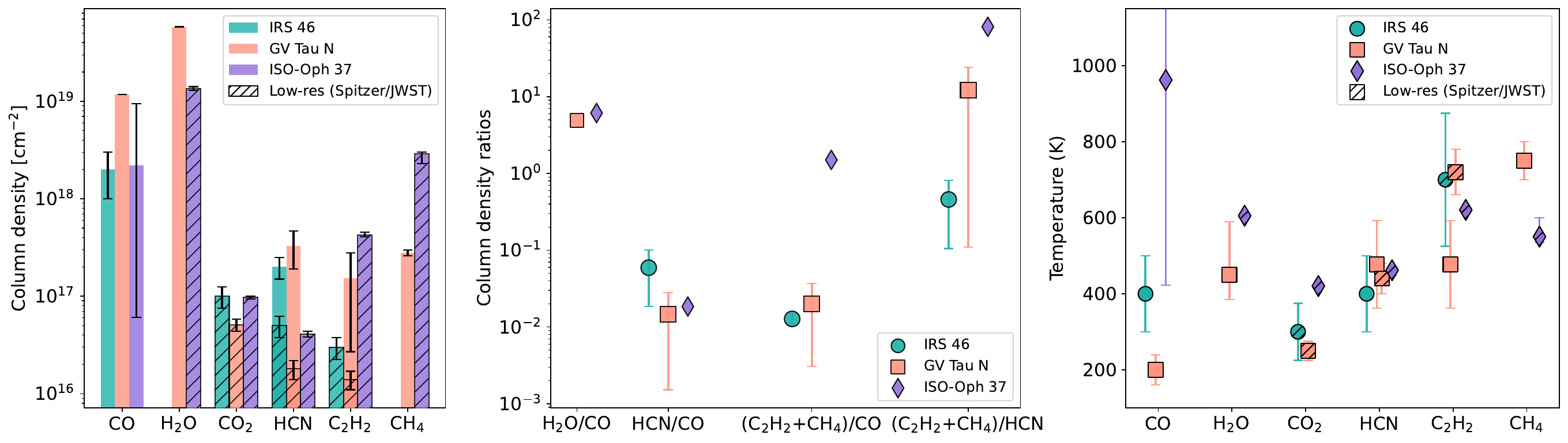}
\caption{Comparison of molecular column densities and temperatures for ISO-Oph~37, IRS~46, and GV~Tau~N. When multiple literature values are available for a given source/species, the absolute column density panel shows the maximum reported value, corresponding to a best-case scenario for hydrocarbon richness. The temperature panel shows the excitation temperatures associated with these maximum column densities. The ranges shown in the column density ratio panel are obtained by combining the minimum and maximum reported column densities for each species, encompassing the diversity of observational constraints. Hatched regions indicate estimates using low-resolution observations. }
\label{fig:source_comparison}
\end{figure*}

To place the results for ISO-Oph~37 in context, we compare our measurements with those of two other systems that exhibit well-characterized molecular absorption: IRS~46 and GV~Tau~N (\autoref{fig:source_comparison}). Both sources are commonly classified as Class~I; however, the absence of extended envelopes and their high inferred inclinations ($\sim75^\circ$ for IRS~46 and $\sim80^\circ$ for GV~Tau~N) have led several studies to argue that they are more consistent with evolved, edge-on Class~II disks \citep{lahuis2006,roccatagliata2011,bast2013,najita2021}. In this respect, these systems are closely analogous to ISO-Oph~37. Literature values for the column densities and excitation temperatures of IRS~46 are drawn from \citet{lahuis2006,bast2013}, while those for GV~Tau~N are compiled from \citet{gibb2007,doppmann2008,gibb2013,bast2013,najita2021}. In \autoref{fig:source_comparison}, we present a comparative view of the molecular column densities, column density ratios, and excitation temperatures for all three sources. We note that these literature estimates are derived from heterogeneous observational data sets using \textit{Spitzer}/IRS, \textit{Keck}/NIRSPEC and \textit{Gemini}/TEXES, spanning spectral resolutions from $R\sim600$ to $100{,}000$. As a result, the comparison is not strictly one-to-one, and systematic differences in wavelength coverage and modeling assumptions (e.g., continuum placement, line lists, line widths, and dilution) can shift inferred column densities and temperatures by factors of a few. Similarly, these systems are known to be variable \citep[e.g.][]{bast2013, najita2021}, hence, the differences in column density might reflect the different observation epochs. When multiple literature values are available for a given species (i.e., HCN, \ce{C2H2}, and \ce{CH4}), the absolute column density panel in \autoref{fig:source_comparison} shows the maximum reported value, corresponding to a best-case scenario for hydrocarbon richness. The temperature panel shows the excitation temperatures associated with these maximum column densities, highlighting the thermal conditions implied by the most hydrocarbon-rich interpretation of each source. In contrast, the column density ratio panel displays the full range of ratios permitted by the literature by combining the minimum and maximum reported column densities for each species, thereby encompassing the range of values inferred from heterogeneous data sets.

For ISO-Oph 37, the columns of HCN and \ce{CO2} are comparable to IRS~46 and GV~Tau N benchmarks (few~$\times ~10^{16}$-$10^{17}$\,cm$^{-2}$), whereas \ce{C2H2} and especially \ce{CH4} are elevated for ISO-Oph 37 by $\sim$0.3-0.7\,dex, pointing to a hydrocarbon-rich line of sight. This is more evident when comparing the \ce{C2H2} columns derived using the same lines (i.e., the $\nu_5$ bending mode at 13.7\,\micron), as the \textit{Spitzer}/IRS estimates for IRS~46 and GV~Tau~N are more than an order of magnitude lower than for ISO-Oph 37. Similarly, when comparing the column density ratios, even in the case where the highest reported \ce{C2H2} and \ce{CH4} columns are assumed, the (\ce{C2H2}$+$\ce{CH4})/HCN values for both GV~Tau~N and IRS~46 are lower than $\sim$25, contrary to the $\sim$80 value found for ISO-Oph 37. This is also the case when comparing the (\ce{C2H2}$+$\ce{CH4})/CO ratio, as it is $\sim$100 times greater for ISO-Oph 37 than for the other two sources. In that context, the higher \ce{C2H2} and \ce{CH4} either reflect an intrinsically carbon-rich chemistry or modest geometry/radiative-transfer effects. A practical caveat is that both \ce{C2H2} and \ce{CH4} have covering fractions $f_c\sim0.5$, whereas the other species have $f_c\sim1$, meaning the latter are directly comparable to other studies, as they typically assumed $f_c=1$; because covering fraction trades off with column density, ratios that mix “partial-covering'' ($f_c=0.5$, \ce{C2H2}, \ce{CH4}) and “full-covering'' ($f_c=1$, HCN, \ce{H2O}) fits should be interpreted with this in mind. We refer the reader to the discussion of the covering fraction covered in \autoref{sec:methods}. Among these organic species, the retrieved temperatures of ISO-Oph 37 are consistent with both sources.

The two most abundant oxygen carriers, CO and \ce{H2O}, have column density estimates that are consistent between ISO-Oph 37 and the other two sources, with CO typically being about a factor of ten ($10^{18}$\,cm$^{-2}$) less abundant than water ($10^{19}$\,cm$^{-2}$), though we note IRS~46 lacks a water measurement. The main difference between sources for these species are the temperatures, with \ce{H2O} in ISO-Oph~37 being slightly hotter (600\,K) than for GV~Tau~N (450\,K), and CO being significantly hotter (1000\,K) than GV~Tau~N (200 K) and IRS~46 (400\, K). As mentioned before, this is likely due to the lines used in the slab model being sensitive to higher temperature gas. 

Overall, the absorption component of ISO-Oph~37 stands out as hydrocarbon-rich, with elevated \ce{C2H2} and \ce{CH4} columns and ratios relative to IRS~46 and GV~Tau~N, despite comparable abundances of the dominant oxygen carriers. These characteristics point to distinct chemical conditions along the line of sight of the absorbing component.

\subsection{Emission}

In addition to the rich absorption spectrum described above, the MIRI/MRS data also reveal emission features that probe a distinct physical and excitation regime. At the shorter-wavelength end of the MIRI/MRS spectrum we detect OH emission (see \autoref{fig:short_lte_fits}). Unlike the nearly symmetric $\Lambda$-doublets expected for thermalized OH, only one component of each doublet is prominent. This is a feature of “prompt” OH produced when \ce{H2O} is photodissociated by FUV photons (often Ly$\alpha$), leaving OH in highly excited, non-thermal states that radiatively decay before collisions can equilibrate the e/f parity levels \citep{carr2014,zhou2015_oh,tabone2024,zannese2024}. Because these lines are demonstrably out of LTE, we do not derive LTE column densities for OH; instead, we compare the observed doublet asymmetries and line ratios to published prompt-OH models \citep{tabone2021,tabone2024}. The presence of prompt OH emission points to gas exposed to an enhanced UV radiation field \citep{neufeld2024}, whose implications for the disk chemistry we discuss in \autoref{sec:discussion}.

Longward of $\sim$15\,\micron\,we also detect two OH rotational components characterized by distinct excitation temperatures ($\sim$500 K and $\sim$1600 K). The warm component is colder than typically seen in T Tauris \citep[e.g.][]{Salyk11,Gasman2023}, but it is similar to some species we see in absorption in this source (e.g., \ce{CO2} and HCN), while the hotter OH likely traces more strongly irradiated or more interior disk layers.

For water, we identify two reservoirs in emission: a warm component at $\sim$440 K and a cooler component at $\sim$200 K, consistent with trends in T Tauri disks from \textit{Spitzer} and JWST surveys \citep{Salyk11,arulanantham2024}. We do not find evidence for a hot ($\geq$800 K) water emission component reported in some systems \citep{banzatti2024,temmink2025,romero-mirza2024}, only the $\sim$600~K absorption component; the absence of hot water emission here could reflect efficient photodissociation in the innermost disk surface. Alternatively, the large inclination (74$^\circ$) could be affecting our ability to trace the innermost disk surface, where the hot water component might reside. 

\subsection{Line shifts}

In previous pre-JWST studies of sources exhibiting molecular absorption, such as GV~Tau~N and IRS~46, velocity shifts relative to the stellar systemic velocity were reported and interpreted as key diagnostics of the absorption origin. Motivated by these results, we assess whether comparable velocity shifts are present in ISO-Oph~37. The details of the line selection used for this analysis, including the specific transitions and selection criteria for each species, are described in Appendix~\ref{app:line_selection}.

\begin{figure}[ht!]

\includegraphics[width=8cm]{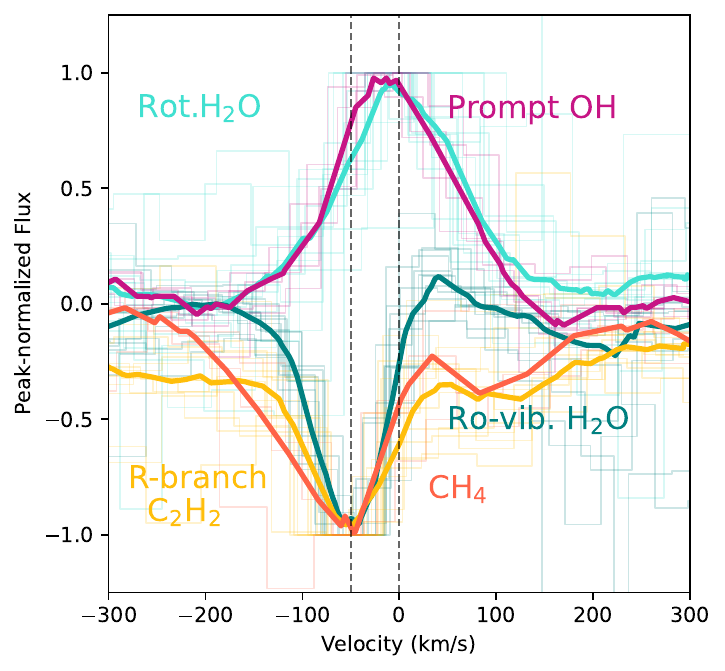}
\caption{Stacked profiles of individual \ce{H2O}, \ce{C2H2}, \ce{CH4} and prompt OH transitions. The lines are in the velocity frame of the source \citep{sullivan2019}. Dashed lines located at 0km/s and -50\,km/s show the blueshift of the absorption component.}
\label{fig:line_shifts}
\end{figure}

Once the lines are selected in the spectrum, we subtract the previously determined continuum and shift the spectrum to the source rest frame using a systemic radial velocity of $-7.9$\,km/s \citep{sullivan2019}. For each selected transition, we convert the spectrum to velocity space around the line center, normalize each line segment by its local peak or trough, and interpolate onto a common velocity grid. We then compute the median of these normalized profiles to construct stacked line profiles for each species, which are used to compare their characteristic line velocity centroids. The individual lines and stacked profiles are shown in \autoref{fig:line_shifts}. From this, it is clear that the central velocities of the \ce{C2H2}, \ce{CH4}, and \ce{H2O} in absorption are blue-shifted up to 50\,km/s with respect to the \ce{H2O} and prompt OH in emission. We note that the smaller absorption peaks that seem to be redshifted, particularly in \ce{CH4}, could be due to contamination from adjacent lines, due to the difficulty of finding isolated line transitions. As noted in Appendix~\ref{app:line_selection}, the \ce{CH4} stacked profile only includes 3 lines. Previous velocity resolution estimates reached with the JDISCS wavelength calibration reported standard deviations of 4.4\,km/s at $<$9\,\micron, 7\,km/s at 9--19\micron, and 10\,km/s at $>$19\,\micron\,\citep{banzatti2024}, confirming our observed shift with $\sim$7$\sigma$ confidence. For the \ce{^12CO} 1-0 data taken with NIRSPEC, the centroids of the fitted Gaussians for the absorption components shown in \autoref{fig:co_slabs} are all blue-shifted, with velocities ranging from 4--15\,km/s, significantly lower than measured in the MIRI-MRS observations.

In IRS~46, both the \ce{^12CO} and HCN lines were reported to be blue-shifted by $\sim$25\,km/s, under the assumption that the source itself is at the cloud velocity \citep{lahuis2006}. These velocities were measured in \textit{Keck}/NIRSPEC observations, which have a velocity resolution of 12\,km/s. In contrast, the \ce{C2H2} and HCN lines of GV Tau N measured with TEXES (3\,km/s resolution), showed red-shifted absorption centered at around 4\,km/s with the red wing of the line extending up to 15--20km/s \citep{najita2021}.

Taken together, the systematic blueshifts observed in the mid-infrared absorption lines, their clear separation from the emission components, and their consistency across multiple molecular species indicate that the absorbing gas is dynamically distinct from both the quiescent disk and the extended cloud environment. The contrast between the large blueshifts measured with MIRI and the more modest shifts seen in the near-infrared CO absorption further suggests that different tracers probe physically and geometrically distinct regions along the line of sight. These results point toward a structured inner-disk environment in which the absorbing material participates in an organized flow rather than purely Keplerian motion, motivating a closer examination of its physical origin in the context of disk winds. We further discuss the implications of these velocity shifts for the origin of the absorbing gas in \autoref{sec:discussion}.

\subsection{Extended emission}

\begin{figure*}[ht!]
\includegraphics[width=18cm]{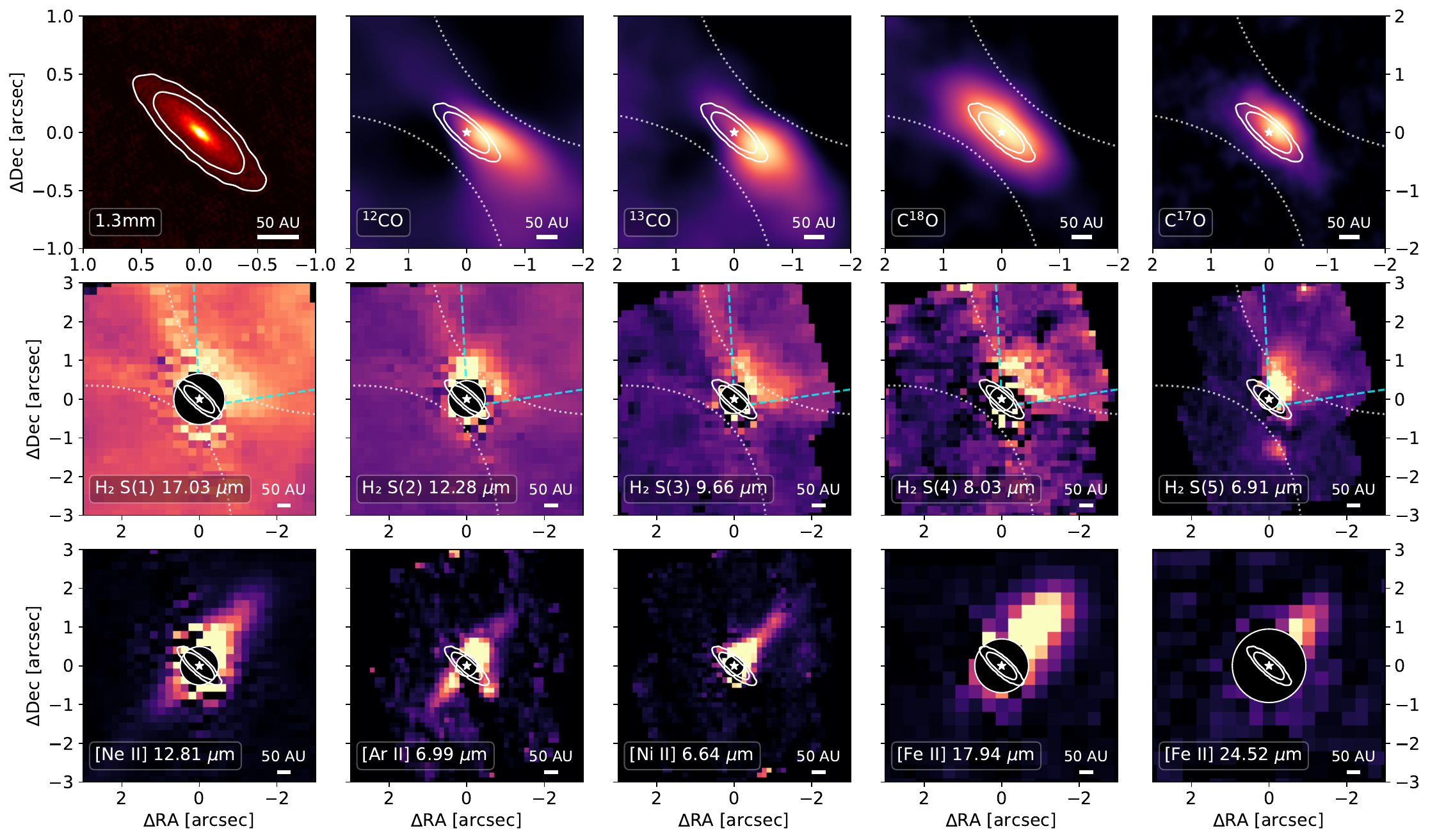}
\caption{ALMA and JWST-MIRI images of ISO-Oph 37. \textit{Top:} ALMA 1.3mm continuum image from ODISEA \citep{cieza2019}, CO isotopologues from AGE-PRO \citep{zhang2025,ruiz-rodriguez2025}. Dotted white lines show the disk surface, traced from the \ce{H2} (S3) line emission from MIRI. \textit{Middle:} \ce{H2} line images from MIRI. The black circle shows the inner working angle of JWST, and the contours show the 1.3mm continuum from ALMA. Blue dashed lines show the disk wind, traced from the \ce{H2} (S5) line emission \textit{Bottom:} atomic line images from MIRI showing jet emission.}
\label{fig:imaging}
\end{figure*}

To characterize the observed geometry of the source, we analyze multi-wavelength images. The source has been observed with ALMA both in ODISEA \citep[as ISO-Oph~37,][]{cieza2019,cieza2021} and in AGE-PRO \citep[as Oph~1/SSTc2dJ162623.6$-$242439,][]{zhang2025,ruiz-rodriguez2025,vioque2025}. In ODISEA, ISO-Oph~37 is classified as a flat-spectrum protostar, and the 1.3\,mm continuum, as shown in \autoref{fig:imaging}, shows a highly inclined ($i=72.4^\circ$), extended, and largely smooth disk at $\sim$4\,au resolution, with only a subtle inflection point at $\sim$31 au \citep{cieza2021}. This morphology is consistent with the broader ODISEA result that embedded and flat-spectrum disks typically lack prominent ring-gap substructure compared to more evolved Class~II disks \citep{cieza2021}. AGE-PRO Band~6 observations recover a consistent inclination ($i=72.6^\circ$), but the improved sensitivity reveals residuals in the inner disk that may be associated with substructure such as a dust trap or a snowline \citep{vioque2025}. In molecular gas,  ISO-Oph 37 exhibits strong contamination of the optically thick $^{12}$CO and $^{13}$CO $J\!\!=\!2\!-\!1$ lines by extended cloud emission associated to the source, whereas the optically thinner C$^{18}$O and C$^{17}$O isolate the Keplerian disk: blue- and red-shifted lobes are recovered cleanly in these isotopologues \citep{ruiz-rodriguez2025}.

To place this emission in context, we generate line images of all pure rotational \ce{H2} and selected atomic lines within the MIRI band. Pure rotational \ce{H2} emission is commonly used to trace extended, low-velocity disk winds \citep[][Narang et al. 2026, submitted]{arulanantham2024, anderson2024, schwarz2025_winds,pascucci2025}, while forbidden atomic lines are established tracers of high-velocity, collimated jets \citep{yang2022,narang2024,vandishoeck2025}, and in some cases also winds \citep{bajaj2024}. The resulting line images, overlaid with continuum contours, are shown in the bottom two rows of \autoref{fig:imaging}. 

The \ce{H2} emission exhibits a complex, excitation-dependent morphology. Lower-excitation transitions show extended emission forming symmetric upper and lower lobes with a convex/concave appearance, most clearly traced by the \ce{H2} S(3) line (outlined by the dotted white curves in \autoref{fig:imaging}). The upper lobe is brighter, consistent with the high inclination of the system and partial obscuration of the far side. The orientation of this extended emission matches that of the C$^{17}$O and C$^{18}$O disk, suggesting that the lower-excitation \ce{H2} primarily traces warm gas along the disk surface.

In contrast, higher-excitation \ce{H2} emission, most notably the S(5) transition, appears more spatially confined and vertically collimated. This morphology closely resembles that observed in other systems hosting disk winds \citep{anderson2024,pascucci2025}, consistent with an origin in hotter gas launched close to the inner disk. The high-excitation \ce{H2} emission is detected only on the upper side of the disk, mirroring asymmetries reported in other wind-hosting disks \citep{bajaj2025}. The atomic lines trace an even more collimated component, oriented perpendicular to the disk. The [Ne~II] and [Ar~II] emission clearly delineate both the northern and southern jets, while [Ni~II] predominantly traces the northern jet. This behavior is consistent with jet emission being stronger along the approaching flow direction and supports a nested morphology in which a fast, collimated jet is embedded within a slower, wider-angle wind, as seen in edge-on systems \citep{pascucci2025}.

\begin{figure*}[ht!]
\includegraphics[width=18cm]{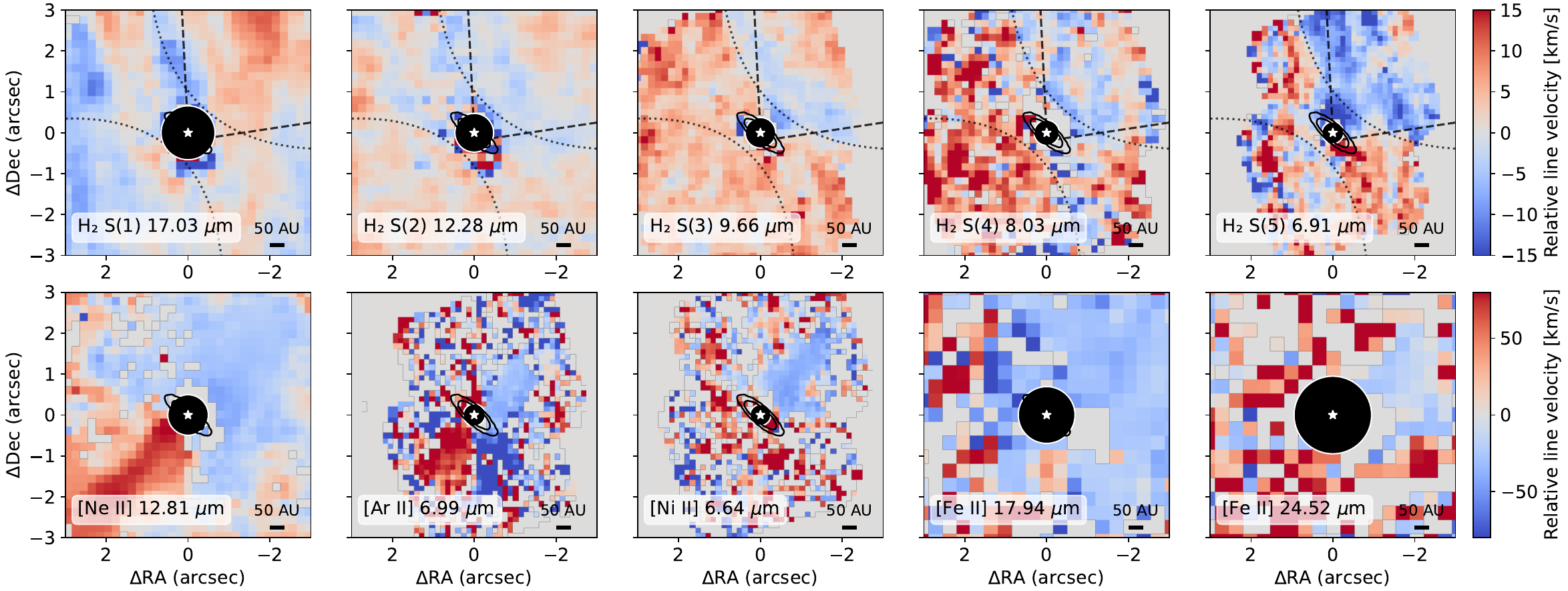}
\caption{Line velocity maps of ISO-Oph 37. The top and bottom panels show the same MIRI lines as \autoref{fig:imaging}. Dotted black lines trace the disk surface inferred from the emission of the \ce{H2} (S3) line, whereas black dashed lines show the disk wind opening angle traced from the \ce{H2} (S5) emission. The solid black line show the 1.3mm continuum from ALMA. The velocities are shown with respect to the source velocity of $v_r=-7.9$\,km\,s$^{-1}$.}
\label{fig:imaging_velocities}
\end{figure*}

To further constrain the origin of the extended emission, we construct velocity maps from the MIRI cubes (\autoref{fig:imaging_velocities}). As expected from the disk inclination and position angle, emission in the upper lobe is predominantly blue-shifted, while the lower lobe is red-shifted, both in \ce{H2} and atomic lines. The \ce{H2} emission associated with the upper-lobe wind exhibits velocities of 5-15\,km\,s$^{-1}$, comparable to the blue-shifted absorption velocities measured in the CO $v=1$-0 line profiles (see \autoref{sec:co_analysis}). This kinematic agreement suggests that the mid-infrared \ce{H2} emission and the CO absorption probe the same wind component. In contrast, the atomic lines trace substantially higher velocities ($\sim$60\,km\,s$^{-1}$), characteristic of jet emission, most clearly seen in [Ne~II] and [Ar~II]. The [Ni~II] and [Fe~II] lines are detected only as blue-shifted emission in the upper lobe, consistent with their spatial distribution in \autoref{fig:imaging}. The presence of blue-shifted CO absorption, extended \ce{H2}, and jets perpendicular to the disk, closely resemble what has been found in other systems such as CX~Tau, where the observed components have been linked to a photoevaporative disk wind product of UV excitation \citep{anderson2024}.

Overall, ISO-Oph~37 presents a dynamically layered environment in which a highly inclined disk is accompanied by both a slow, warm disk wind and a faster, collimated jet. The clear stratification in excitation, morphology, and kinematics across molecular and atomic tracers points to a nested structure linking the disk surface to progressively more energetic outflow components.

\section{Discussion} \label{sec:discussion}

\begin{figure*}[ht!]
\includegraphics[width=18cm]{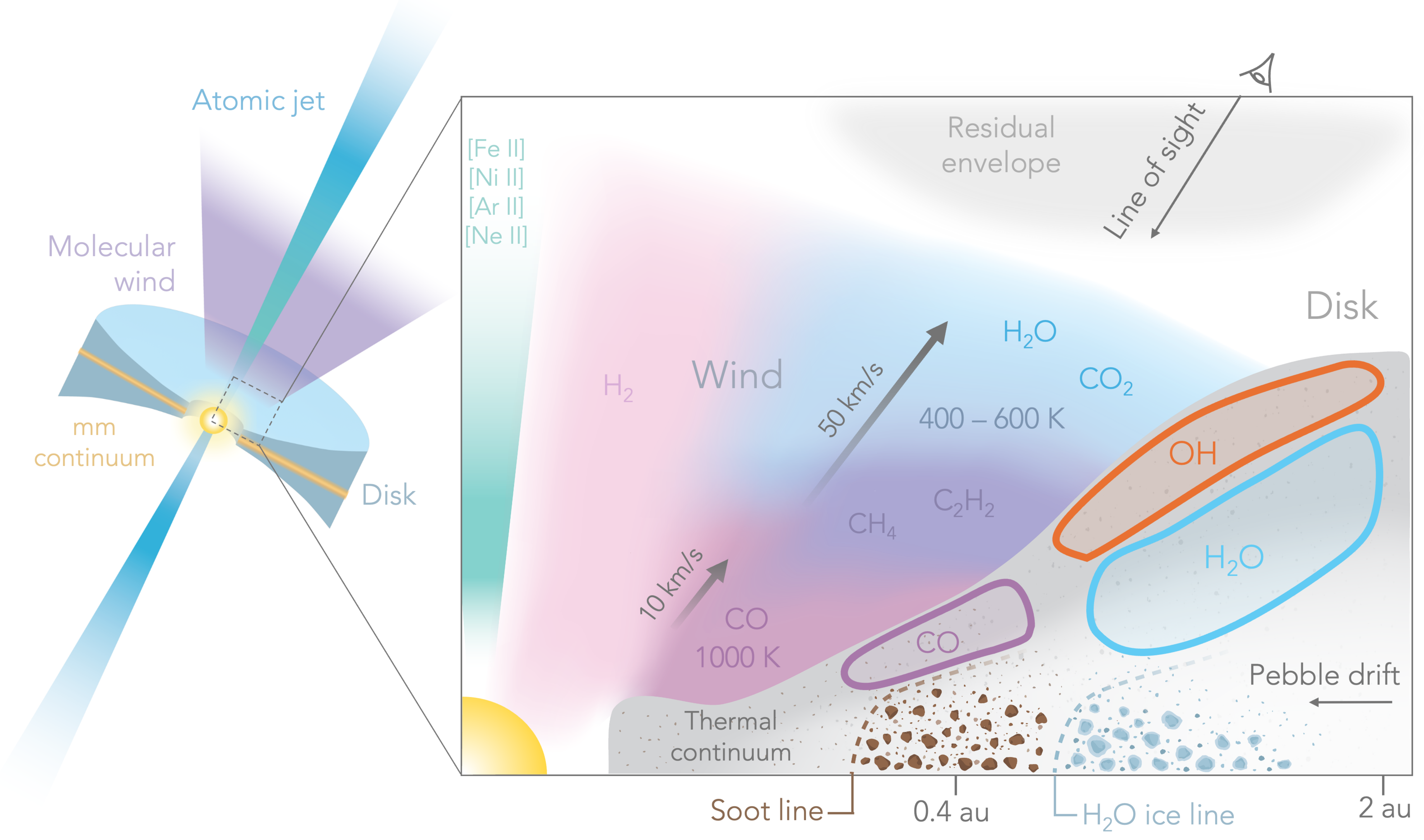}
\caption{Schematic diagram of ISO-Oph 37. The thermal continuum serves as a background for some of the species in the wind (CO, \ce{C2H2}, \ce{CH4}, \ce{H2O}, \ce{CO2}) to absorb against. The thick arrows show the direction of the wind velocity. The molecular labels in the wind are placed to indicate relative velocity, temperature, and covering fraction, and should not be interpreted as a chemical progression along the wind. All species are expected to originate in the inner disk atmosphere and be entrained in the outflow. Solid contours show the species detected in emission.}
\label{fig:masterpiece}
\end{figure*}

Based on the results presented in the previous sections, we outline a schematic representation of our interpretation of the relative locations of the emission and absorption components observed in ISO-Oph~37. This schematic, shown in \autoref{fig:masterpiece}, serves as a visual guide and is referenced throughout the discussion below. We note that the placement of molecular species along the wind is purely schematic and is intended to illustrate their inferred velocities, excitation temperatures, and covering fractions. It should not be interpreted as a chemical sequence or as implying in-situ synthesis along the flow. Instead, all species are expected to be present in the inner disk atmosphere and to be entrained in the wind, with their apparent spatial separation reflecting differences in kinematics and excitation rather than chemical evolution.

\subsection{Origin of the absorption}
\label{subsec:origin_absorption}

The excitation conditions and column densities of the absorbing gas strongly argue that the mid-IR absorption arises close to the inner disk rather than in the foreground cloud or a cold outer envelope. The hydrocarbons, \ce{CO2}, and HCN exhibit excitation temperatures of $T_{\rm ex}\sim 400$-600\,K, hotter than what is typically found in the envelopes of more massive YSOs \citep{an2011}, and high columns (log $N$/cm$^{2}\sim 10^{16}$-$10^{19}$), far exceeding what is expected for quiescent cloud material. The absorbing layer must also lie very close to the continuum source; most species have covering fractions consistent with unity, while \ce{C2H2} and \ce{CH4} require $f_{\rm c}\approx 0.6$, suggestive of a partially filled line of sight where warm gas covers much, but not all, of the mid-IR continuum-emitting surface. In a highly inclined system ($i\sim74^\circ$), such partial covering naturally arises if the absorption originates in a geometrically thin but vertically extended structure, such as the base of a disk wind, rather than in a static, axisymmetric disk atmosphere.

Similar excitation conditions and molecular inventories have been reported toward IRS~46 and GV~Tau, as both systems show mid-IR absorption from hot ($\gtrsim$350-700\,K) organic gas located within the inner few au and viewed at high inclination. In those systems, the absorbing material has been interpreted as originating either from the base of a disk wind (in the case of the blueshifted absorption in IRS~46; \citealt{lahuis2006}) or from accretion flows along the disk surface (in the case of the redshifted absorption in GV~Tau; \citealt{najita2021}). These phenomena may represent different manifestations of the same underlying magnetically driven flow structure, as global non-ideal MHD simulations \citep[e.g.,][]{bai2017} show that the inner disk can simultaneously host wind-driven outflows and surface accretion streams at different heights, with asymmetric magnetic field configurations naturally producing spatially distinct inflow and outflow components. This very dynamic environment is also consistent with the fact that the sources have shown variability in timescales of years \citep{bast2013,najita2021}.

In the case of ISO-Oph 37, we support a disk wind origin with a velocity- and temperature-stratified structure, based on the kinematic information. The MIRI \ce{C2H2} and \ce{H2O} absorption bands are blueshifted by 50\,km\,s$^{-1}$ relative to the systemic velocity, whereas the CO fundamental absorption observed with \textit{Keck}/NIRSPEC is centered at only $-10$\,km\,s$^{-1}$ in the same frame. Such large velocity offsets are inconsistent with foreground cloud material in $\rho$~Oph but are comparable to the blueshifts observed in near-IR molecular absorption toward IRS~46 \citep{lahuis2006}.

Rather than indicating distinct physical origins for the different molecular species, we interpret these velocity differences as signatures of acceleration within a molecular disk wind. Gas launched near the base of the flow is expected to accelerate outward, so that material farther from the star can reach higher outflow velocities while cooling. The MIRI-traced species would then probe cooler, faster gas, whereas the CO absorption traced by the NIRSPEC lines reflects a hotter gas closer to the launching region that has not yet accelerated to the higher velocities seen at larger distances, as represented in \autoref{fig:masterpiece}. The presence of these organic species in a wind structure is also supported by the wind emission clearly observed in the \ce{H2} lines.

The interpretation of a disk wind origin is consistent with magnetohydrodynamic (MHD) disk wind models, which predict acceleration and thermal stratification along the outflow \citep{shang2023,tu2025}. The observed association of higher velocities with lower excitation temperatures in ISO-Oph~37 can therefore be explained within a single accelerating MHD wind, without invoking multiple unrelated components. Although the precise launch region of the outflow remains uncertain, its close spatial and kinematic association with the disk suggests that the absorbing gas is drawn from, or strongly coupled to, disk material. The molecular composition of the wind is therefore expected to broadly reflect that of the disk, rather than originating in a chemically distinct envelope component.

\subsection{Evolutionary stage}

Several independent pieces of evidence point to ISO-Oph~37 being in a transitional evolutionary stage between an embedded protostar and a more settled Class~II disk. Its low bolometric temperature ($\sim$670\,K) place it right at the edge of Class I/II classification \citep{hsieh2025}. Similarly, its SED has long been classified as flat-spectrum or borderline Class~II \citep{greene1994,McClure10,ruiz-rodriguez2025,vioque2025}, and the new MIRI spectrum presented here confirms a strong, rising mid-IR continuum attenuated by deep silicate features, indicative of substantial line-of-sight extinction. Nonetheless, it is known that high inclinations can make Class II objects appear as if they have an additional far infrared excess from an envelope \citep{robitaille2006}. Due to this degeneracy, we also explored the possibility of the inclination being the source of excess rather than the residual envelope. We modeled the full SED of the source, shown in \autoref{fig:full_spec}, using a grid of 180,000 synthetic SEDs designed for young stellar objects \citep{robitaille2017,richardson2024}. These models cover a wide range of evolutionary stages by assuming different combinations of components (e.g. star, disk, envelope). The details of the modeling are presented in \autoref{app:sed_modeling}; here, we only discuss the implications of the results, noting that the modeling is only meant to explain the overall behavior of the system rather than to find the ``best-fit" parameters. 

In the SED modeling, we find that the solutions that most closely resemble the observed SED do not uniquely prefer a purely disk-only solution: some models including a modest envelope and high viewing inclinations reproduce the broadband SED about as well as disk-only configurations, also with high inclinations. This degeneracy between geometry (edge-on vs.\ face-on) and evolution (embedded vs.\ Class~II) is a well-known limitation of SED-based classifications, especially in high-extinction regions such as $\rho$~Oph where the mid-IR extinction law deviates from the ISM and can bias spectral indices \citep{McClure2009}. This highlights the ambiguity of spectral index classifications, and the need to interpret the system integrating multiple observations. 

In this context, the extended cloud emission detected in \ce{^12CO} and \ce{^13CO} may trace a low-mass remnant envelope associated with the early formation of the system \citep{ruiz-rodriguez2025}, pointing to a dynamically complex environment around ISO-Oph~37. Consistent with this picture, the AGEPRO survey measured a dust radius of $R_{95\%}=104\,\rm au$ \citep{vioque2025}, significantly larger than is typically observed for disks in Ophiuchus and notably lacking clear substructures \citep{cieza2019}. In addition, ISO-Oph~37 hosts the second-highest gas mass in the AGEPRO sample, with $M_{\rm gas}=0.23\,M_\odot$ \citep{trapman2025_gasmass}. Taken together, these properties suggest that ISO-Oph~37 harbors a relatively massive disk that may still be embedded in, or partially surrounded by, a low-mass remnant envelope, rather than representing a fully exposed, evolved T~Tauri disk. Another possibility is that the extended emission from \ce{^{12}CO} and \ce{^{13}CO} is tracing a large-scale streamer interacting with the disk, as has been proposed to explain the mid-IR emission in other systems \citep{perotti2026}. However, the current data do not provide sufficient spatial or kinematic constraints to assess this scenario, and further observations of multiple molecular tracers on larger scales would be required to investigate this possibility.

At the same time, the molecular inventory of ISO-Oph~37 contrasts sharply with that of other young and embedded sources observed with JWST. While deeply embedded or edge-on systems at similar evolutionary stages commonly exhibit absorption dominated by ices formed at large radii \citep[e.g.,][]{sturm2024}, the MIRI spectrum of ISO-Oph~37 is instead dominated by warm gas-phase absorption from \ce{C2H2}, \ce{HCN}, \ce{CO2}, \ce{CH4}, and \ce{H2O}, with excitation temperatures of $T_{\rm ex}\sim400$-600\,K. This places ISO-Oph~37 among the small subset of sources in which absorption traces warm inner-disk gas rather than cold envelope material, such as IRS~46 and GV~Tau~N \citep{bast2013,najita2021}. 

However, even when compared to these sources, its spectrum is particularly rich in hydrocarbons, with prominent \ce{C2H2} and \ce{CH4} absorption, pointing to carbon-rich conditions in the inner few au. Although the absorbing gas is likely observed in an outflowing configuration, it must ultimately originate from disk material. The molecular absorption therefore reflects the composition of the inner disk at the point where gas is lifted into the wind, rather than that of a chemically distinct envelope. The presence of such a molecular inventory in a source that otherwise appears young and partially embedded indicates that substantial chemical processing of disk gas can occur early, before the system has fully evolved into a Class~II disk, emphasizing that carbon-rich conditions might not be exclusive to system in later stages of their evolution \citep{long2025,arabhavi2025}.

\subsection{Presence and origin of excess carbon}

The unusually high inferred (\ce{C2H2}+\ce{CH4})/HCN column density ratio in ISO-Oph~37 — and to a lesser extent the (\ce{C2H2}+\ce{CH4})/CO ratio given the differences in excitation temperature—can be understood in at least two limiting scenarios. In the first scenario, the chemistry would be regulated by an increased UV radiation field; CO would be efficiently photodissociated in the presence of very strong UV, and the liberated carbon is subsequently processed into hydrocarbons through ion-molecule and high-temperature neutral-neutral chemistry, boosting the (\ce{C2H2}+\ce{CH4})/CO ratio. This scenario would be partially supported by the presence of prompt OH emission in the MIRI spectrum, as it requires \ce{H2O} photodissociation. Nonetheless, both the absorption and emission components of \ce{H2O} still contain high columns of gas (see \autoref{tab:slab_vals}), indicating that the liberated oxygen might be reacting to reform \ce{H2O} faster than the FUV can photodissociate it. Similarly, the large \ce{H2O} columns inferred in both absorption would likely provide substantial self-shielding and mutual shielding of the wind material \citep{Bethell09}, attenuating the FUV radiation field and making it difficult for UV photons to penetrate deeply enough to significantly deplete CO.

An alternative interpretation is that the CO and HCN abundances are typical for warm gas close to the inner disk, as suggested by the HCN/CO ratios, which are comparable to those measured in other sources (\autoref{fig:source_comparison}), while the hydrocarbons, and to a lesser extent water, are enhanced. The expectation is that if most of the volatile oxygen is stored in water, the abundances reflect \ce{H2O}/CO$\approx$1.4-2 \citep{vanDishoeck2021}. This makes the \ce{H2O}/CO$\approx$6 found in ISO-Oph~37 unusually high, although in line with the estimate for GV~Tau~N (see \autoref{fig:source_comparison}). This high \ce{H2O}/CO ratio could then be driven by an overabundance of water, which can be explained by a supply of \ce{H2O} icy pebbles \citep{krijt2025}. Water abundances in the inner disk are expected to peak between $0.5-1$\,Myr due to the sublimation of water ice around inward drifting pebbles \citep{kalyaan2021,Kalyaan23, houge2025_water}. Based on the relative incidence of different SED classes in nearby molecular clouds, \citet{Evans09} estimated a statistical age of 0.9~Myr for the flat-spectrum phase, making the age of ISO-Oph 37 roughly correspond to the peak water abundance in the inner disk. In addition, the lack of gaps in the millimeter emission \citep{cieza2021,vioque2025} suggests that there are no strong substructures or dust traps halting the drift of particles.

At the same time, the excess \ce{C2H2} and \ce{CH4} could arise from the sublimation and thermal processing of carbonaceous grains at the soot line, where refractory carbonaceous material is destroyed, releasing volatile carbon into the gas phase \citep{Li21,colmenares2024b,houge2025_carbon}. In this second scenario, the ratios we measure are signaling early and efficient redistribution of \emph{both} oxygen and carbon reservoirs by pebble drift and soot-line chemistry, rather than a deficit of CO.

As discussed before, the differences in spectral resolution and coverage can affect the column densities estimated for each absorption system. Nonetheless, the HCN and \ce{C2H2} columns shown in \autoref{fig:source_comparison} that were inferred using the same wavelength coverage with IRS and MIRI-MRS, and that have similar temperatures across different systems, still point towards an excess of carbon in ISO-Oph 37, regardless of the CO content. This result then favors the carbon enhancement scenario due to the sublimation of refractory carbon grains. It is important to note that the excess carbon does not necessarily imply a high C/O ratio, as water is also found in high abundances. This is consistent with thermochemical models where excess oxygen from destruction of CO is driven to \ce{H2O} faster than CO can re-form, and hence the excess carbon produced from grain sublimation does not result in an elevated C/O ratio despite the extra carbon being locked in hydrocarbon-chains \citep{duval2022}. 

More generally, this interpretation then has direct implications for the carbon reservoir available to forming planets. If refractory carbon is retained in solids, it can lead to the formation of carbon-rich (``soot-rich") planets \citep{li2026_sootworld}. On the contrary, if a significant fraction of the carbon in disks is released to the gas phase and subsequently entrained in a disk wind, then the inner disk may become depleted in solid-phase carbon before it can be incorporated into planetesimals, favoring carbon-poor terrestrial compositions. In this scenario, the same processes that lead to the observed hydrocarbon enhancements in the gas are simultaneously reducing the carbon available to build rocky planets. This system then provides a link between early pebble drift and carbon grain processing, to the carbon-poor composition of terrestrial planets like Earth, relative to their natal disk environments \citep{Li21}.

\section{Conclusions} \label{sec:conclusion}

We presented JWST/MIRI-MRS and \textit{Keck}/NIRSPEC observations of ISO-Oph~37, a flat-spectrum source in the $\rho$~Ophiuchi star-forming region. The spectrum reveals a combination of mid-infrared molecular absorption (CO, \ce{C2H2}, \ce{CH4}, HCN, \ce{CO2}, and \ce{H2O}) and emission from \ce{CO}, \ce{H2O}, and OH. By combining LTE slab modeling, spatially resolved line imaging, kinematic analysis, and SED modeling, we arrive at the following conclusions:

\begin{enumerate}

\item ISO-Oph~37 is best interpreted as a transitional system between an embedded protostar and a Class~II disk. Its flat-spectrum SED, high inclination, smooth but moderately extended millimeter continuum, and large inferred gas mass are consistent with a massive disk that may still be partially embedded in, or interacting with, a low-mass remnant envelope. This evolutionary ambiguity highlights the limitations of SED-based classifications in highly inclined systems.

\item The JWST/MIRI spectrum reveals a rich molecular inventory that differs markedly between emission and absorption. Emission from \ce{H2O} and OH traces warm disk surface layers exposed to strong FUV irradiation, while the absorption spectrum probes warm gas close to the inner disk with $T_{\rm ex}\sim400$-600~K, dominated by \ce{C2H2}, \ce{CH4}, HCN, \ce{CO2}, and \ce{H2O}. The high column densities inferred from absorption indicate that this gas lies close to the mid-infrared continuum source, within the inner few astronomical units.

\item The molecular absorption is kinematically distinct from the quiescent disk. The large blueshifts observed in the mid-infrared absorption lines (tens of km\,s$^{-1}$), together with the more modest blueshifts detected in near-infrared CO absorption, rule out a static foreground or envelope origin. Instead, the velocity structure is consistent with a velocity- and temperature-stratified molecular disk wind, in which material launched from the inner disk accelerates outward. In this framework, the absorbing gas preserves the chemical imprint of the disk at the wind-launching region.

\item Compared to the benchmark absorption systems IRS~46 and GV~Tau~N, ISO-Oph~37 exhibits systematically enhanced \ce{C2H2} and \ce{CH4} column densities and unusually high (\ce{C2H2}+\ce{CH4})/HCN and (\ce{C2H2}+\ce{CH4})/CO column density ratios. In contrast, the column densities of CO and HCN are comparable to those measured in the other absorption systems. These ratios therefore primarily reflect an overabundance of hydrocarbons rather than a deficit of CO and HCN.

\item The observed molecular ratios are most naturally explained by an enhancement of the inner-disk carbon reservoir rather than by UV-driven CO destruction. Inward drift and sublimation of icy pebbles can deliver excess \ce{H2O} to the inner disk, while sublimation and thermal processing of refractory carbonaceous grains at the soot line can efficiently recycle solid carbon into volatile hydrocarbons such as \ce{C2H2} and \ce{CH4}. In this scenario, the molecular disk wind acts as a tracer of disk chemistry, carrying an imprint of early pebble-driven redistribution of carbon and oxygen into the observed absorption spectrum.

\end{enumerate}

ISO-Oph~37 thus demonstrates that carbon-rich inner-disk chemistry associated with pebble drift and soot-line processing can be established at relatively early evolutionary stages ($\lesssim$1~Myr) and be directly sampled through molecular absorption in disk winds. This loss of carbon-rich gas through disk winds may reduce the amount of carbon locked into solids in the inner disk, thereby affecting the composition of forming terrestrial planets. Absorption-selected systems like ISO-Oph~37 therefore provide powerful laboratories for linking disk chemistry to outflowing gas and for constraining how solid-phase carbon is recycled into the gas phase in planet-forming regions, a process that JWST is now uniquely positioned to explore across larger samples.

\begin{acknowledgments}

The authors thank the anonymous reviewer for their suggestions, which helped improve the quality of the paper. M.J.C. thanks Nuria Calvet, Lee Hartmann, and August Masley for helpful discussions and guidance regarding the SED modeling presented in this work. SK and TK acknowledge support from STFC Grant ST/Y002415/1.  L.C. acknowledges support from ANID -- Millennium Science Initiative Program -- Center Code NCN2024\_001. A portion of this research was carried out at the Jet Propulsion Laboratory, California Institute of Technology, under a contract with the National Aeronautics and Space Administration (80NM0018D0004). This work is based on observations made with the NASA/ESA/CSA James Webb Space Telescope. The data were obtained from the Mikulski Archive for Space Telescopes at the Space Telescope Science Institute, which is operated by the Association of Universities for Research in Astronomy, Inc., under NASA contract NAS 5-03127 for JWST. The JWST data is available at MAST: \dataset[doi: 10.17909/h1me-fk19]{\doi{10.17909/h1me-fk19}.} The W. M. Keck Observatory was made possible by the generous financial support of the W. M. Keck Foundation. We would like to thank all Keck Observatory staff who facilitated our observations. We recognize the cultural significance that the summit of Maunakea has within the indigenous Hawaiian community. We are deeply grateful for the opportunity to conduct observations from this mountain, while acknowledging the impact of our presence and the ongoing efforts to preserve this culturally and environmentally significant site. This paper makes use of the following ALMA data:ADS/JAO.ALMA\#2021.1.00128.L. ALMA is a partnership of ESO (representing its member states), NSF (USA) and NINS (Japan), together with NRC (Canada), MOST and ASIAA (Taiwan), and KASI (Republic of Korea), in cooperation with the Republic of Chile. The Joint ALMA Observatory is operatedby ESO,AUI/NRAO and NAOJ. The National Radio Astronomy Observatory is a facility of the National Science Foundation operated under cooperative agreement by Associated Universities, Inc. We acknowledge the use of ChatGPT to assist with debugging code in the MCMC implementation and for minor language revisions to improve clarity and readability. All results were independently verified by the authors, and all scientific analyses and interpretations were conducted by the authors. The manuscript further benefited from language editing by all co-authors.

\end{acknowledgments}

\facilities{JWST (MIRI), Keck (NIRSPEC), ALMA.}

\software{\texttt{numpy} \citep{numpy_2020}, \texttt{scipy} \citep{scipy2020}, \texttt{Astropy} \citep{Astropy22}, \texttt{spectools-ir} \citep{spectools2022}, \texttt{emcee} \citep{Foreman13}, \texttt{sedfitter} \citep{robitaille2017, robitaille2017zndo}, \texttt{Chat-GPT} \citep{chatgpt2026}.}

\appendix

\section{MCMC modeling details}
\label{app:priors}

\subsection{Emission}

For the emission modeling, we adopt uniform priors on all slab parameters within physically motivated bounds, guided by previous analyses of mid-infrared molecular emission in disks \citep[e.g.,][]{arulanantham2025}. The free parameters for each emitting component are the excitation temperature $T_{\rm ex}$, column density $N_{\rm col}$, and emitting area $A_{\rm em}$. The adopted prior ranges are listed in \autoref{tab:priors}. To stabilize the decomposition of multiple water emission components, we impose an additional Gaussian prior on the excitation temperature of the cold water component, centered at $T_{\rm ex}=200$\,K with a standard deviation of $\sigma=50$\,K. 

In addition, to promote rapid convergence of the MCMC sampling, walkers are initialized near representative starting values, listed as $T_{\rm ini}$, $\log N_{\rm ini}$, and $\log A_{\rm ini}$ in \autoref{tab:priors}. These values are obtained by iteratively adjusting the parameters until the corresponding slab model provides a reasonable visual match to the observed spectrum. Importantly, these initial conditions are used solely to initialize the chains and do not affect the resulting posterior distributions, which are determined entirely by the likelihood function and the adopted priors.

\begin{deluxetable*}{lcccc}[ht!]
\tablecaption{Priors adopted for the emission slab-model MCMC fits.\label{tab:priors}}
\tablehead{
\colhead{Parameter} &
\colhead{\ce{H2O}$_{\rm warm}$} &
\colhead{\ce{H2O}$_{\rm cold}$} &
\colhead{\ce{OH}$_{\rm warm}$} &
\colhead{\ce{OH}$_{\rm hot}$}
}
\startdata
$T_{\rm ex,ini}$ [K]        & 450  & 200 & 400 & 900 \\
$T_{\rm ex,min}$ [K]        & 200  & 100 & 100 & 100 \\
$T_{\rm ex,max}$ [K]        & 800  & 400 & 1000 & 3000 \\
\hline
$\log N_{\rm col,ini}$ [cm$^{-2}$] & 18.2  & 17.0 & 17.5 & 16.0 \\
$\log N_{\rm col,min}$ [cm$^{-2}$] & 12.0  & 12.0 & 12.0 & 12.0 \\
$\log N_{\rm col,max}$ [cm$^{-2}$] & 22.0  & 22.0 & 22.0 & 22.0 \\
\hline
$\log A_{\rm slab,ini}$ [au$^{2}$] & 0.75  & 2.0  & 0.6  & 0.5 \\
$\log A_{\rm slab,min}$ [au$^{2}$] & -2.0  & -2.0 & -2.0 & -2.0 \\
$\log A_{\rm slab,max}$ [au$^{2}$] & 2.0   & 2.5  & 2.0  & 2.0 \\
\enddata
\end{deluxetable*}

\subsection{Absorption}

For the absorption modeling, we adopt uniform priors on the excitation temperature, column density, and covering fraction of each absorbing species:
\begin{align*}
T_{\rm ex} &\sim \mathcal{U}(100,1500)\ {\rm K},\\
\log N_{\rm col} &\sim \mathcal{U}(16,26)\ \ [{\rm cm^{-2}}],\\
f_c &\sim \mathcal{U}(0,1].
\end{align*}

In the 12--16\,\micron\ interval affected by strong silicate absorption and potential hydrocarbon pseudo-continuum, the continuum is modeled simultaneously with the molecular absorption using a cubic spline. The spline is evaluated at a fixed set of node wavelengths $\lambda_j$, while the corresponding node fluxes $F_{c,j}$ are treated as free parameters with uniform priors,
\begin{equation}
F_{c,j} \sim \mathcal{U}(0.5,3.0)\ {\rm Jy}, \qquad j=1,\dots,8,
\end{equation}
ensuring positive and physically plausible continuum levels. We note that the continuum flux values $F_{c,j}$, as well as the best-fit continuum fluxes reported in \autoref{tab:spline_nodes}, correspond to the de-reddened flux levels. The spline-defined continuum is multiplied by the absorption model before comparison with the observed spectrum.

For isotopologues expected to trace the same absorbing gas reservoir, additional coupling constraints are imposed. In particular, the excitation temperature and covering fraction of \ce{^{13}CCH2} are required to remain within 10\% of the corresponding \ce{C2H2} values,
\begin{align*}
\left|T_{\rm ex}(\ce{^{13}CCH2})-T_{\rm ex}(\ce{C2H2})\right| &\le 0.1\,T_{\rm ex}(\ce{C2H2}),\\
\left|f_c(\ce{^{13}CCH2})-f_c(\ce{C2H2})\right| &\le 0.1,
\end{align*}
reflecting the assumption that both species arise from the same absorbing layer.

\begin{deluxetable}{cc}[ht!]
\tablecaption{Spline node wavelengths and best-fit continuum fluxes (before extinction) for the absorption modeling in the 12--16\,\micron\ region.\label{tab:spline_nodes}}
\tablehead{
\colhead{$\lambda_j$ [$\micron$]} &
\colhead{$F_{c,j}$ [Jy]}
}
\startdata
12.0011 & 2.1747 \\
12.5722 & 2.3143 \\
13.1433 & 2.3248 \\
13.7143 & 2.3701 \\
14.2854 & 2.4441 \\
14.8565 & 2.4433 \\
15.4276 & 2.4518 \\
15.9987 & 2.4992 \\
\enddata
\end{deluxetable}

We include the posterior distributions for the CO slab modeling from \autoref{sec:co_analysis}, showing that choosing the MAP over the median value for the CO column density results in an increase by a factor of $\sim$2 in column density. We emphasize that this result would not affect our overall conclusions, as the decrease in CO column density would only increase (\ce{C2H2}+\ce{CH4})/CO ratio found for ISO-Oph 37, making it even more extreme than GV~Tau~N and IRS~46 (see \autoref{fig:source_comparison}).

\begin{figure}[ht!]
\includegraphics[width=8cm]{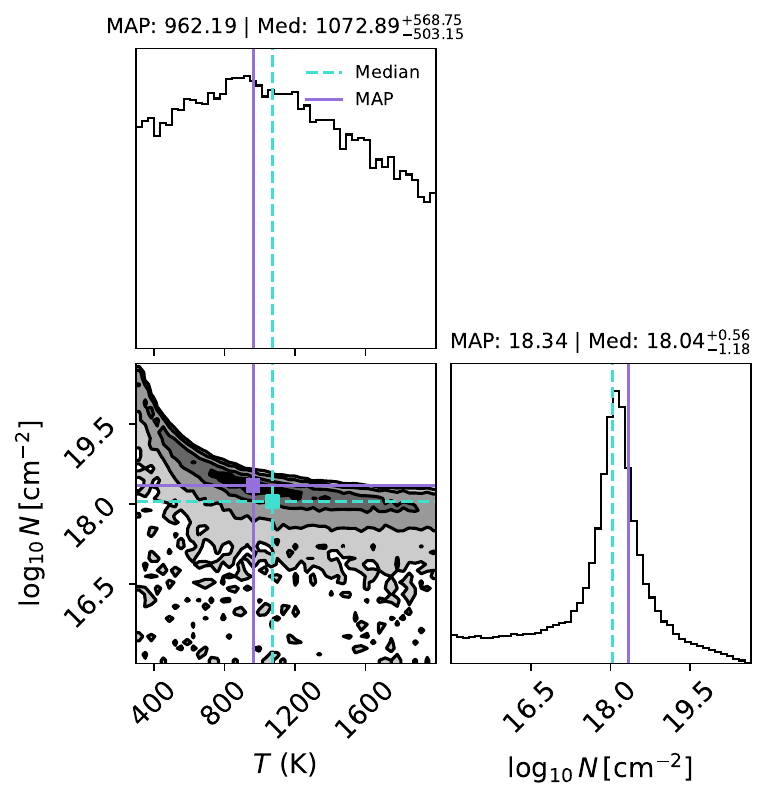}
\caption{Posterior distribution from the MCMC slab model fitting for the CO $v=1$-- absorption component.}
\label{fig:co_posterior}
\end{figure}

\section{\ce{^{13}CO} in \textit{Keck}/NIRSPEC data}
\label{app:13co}
\begin{figure}[ht!]
\includegraphics[width=8.5cm]{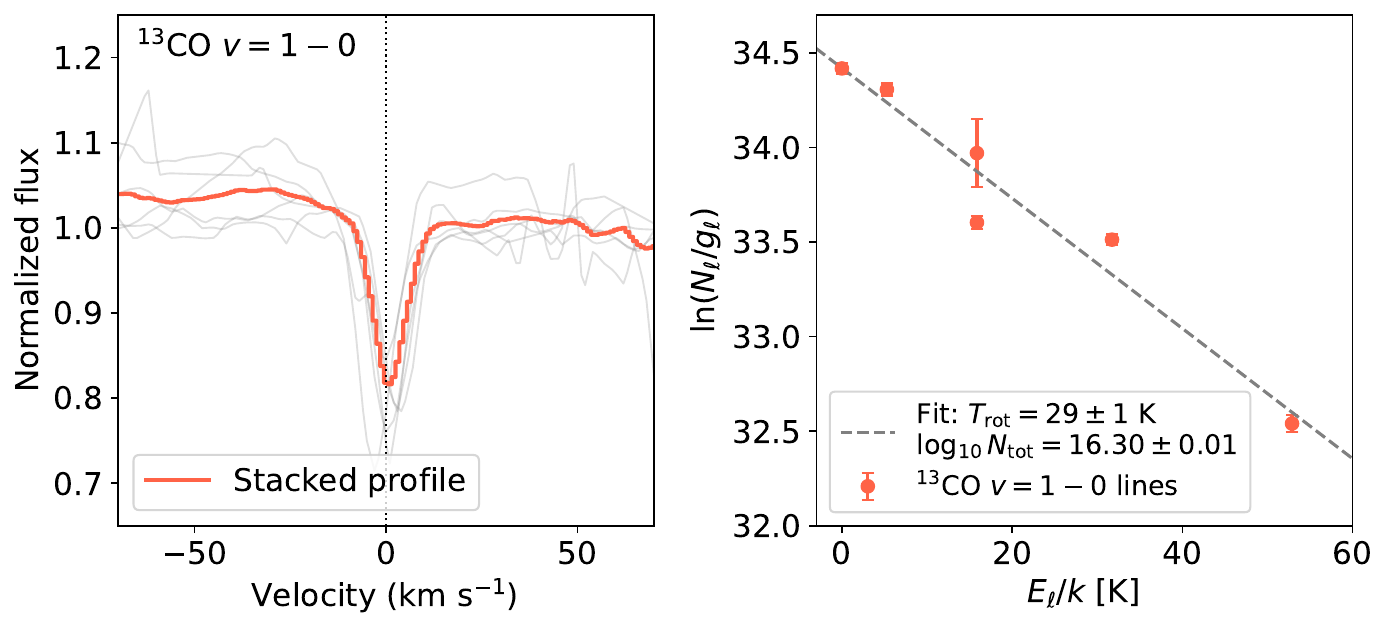}
\caption{\textit{Left:} \ce{^{13}CO} stacked profile. Grey lines show individual line transitions, vertical dotted line shows rest velocity. \textit{Right:} rotation diagram of \ce{^{13}CO} line absorption. }
\label{fig:13co}
\end{figure}

In addition to the \ce{^{12}CO} transitions detected in the NIRSPEC spectrum, we also detect the presence of \ce{^{13}CO}. To characterize the physical conditions of this gas, we perform a rotation diagram of all detected transitions that are separated by at least 10\,km\,s$^{-1}$ from a \ce{^{12}CO} line. Since the detected lines are only in absorption, we follow the methodology of previous works \citep[e.g.,][]{li2022_co} to derive a column density and an excitation temperature from the observed optical depth. We also build a stacked profile of the transitions to asses for bulk velocity shifts, following the procedure outlined in \autoref{sec:co_analysis}. We show the stacked profile, made up of transitions R(0), R(1), and P(1)-P(4) in \autoref{fig:13co}. Overall, the \ce{^{13}CO} lines seem to be tracing a much colder ($T_{\rm ex}\sim 30$\,K) component, likely from the extended envelope seen in the ALMA images. This is also in agreement with the lack of a velocity shift, as seen in the stacked profile, and the fact that the line is seen purely in absorption. 

\section{Line Selection for Velocity Shift Analysis}
\label{app:line_selection}
To search for velocity shifts in the molecular absorption and emission lines, we selected a set of unblended or minimally blended transitions for each species. For \ce{H2O}, we adopted unblended rovibrational and rotational transitions within the MIRI wavelength range from \citet{banzatti2024}. For \ce{C2H2}, the absorption is dominated by relatively isolated R-branch features (see \autoref{fig:lte_fits}). We therefore selected R-branch transitions from the HITRAN database in the 12.9-13.4\,\micron\ range with line intensities exceeding $10^{-19}\,\mathrm{cm^{-1}\,(molecule\,cm^{-2})^{-1}}$, yielding a total of 12 transitions. For absorption features, the HITRAN line intensity provides an appropriate metric for line selection, as it directly reflects the expected absorption strength under LTE conditions. For \ce{CH4}, which lies within a dense forest of \ce{H2O} absorption lines, we first identified wavelength regions where \ce{CH4} transitions are relatively isolated from nearby \ce{H2O} features. Within these regions, we selected the transitions with the highest line intensities, resulting in three usable \ce{CH4} lines.
Finally, for the prompt OH emission, we searched the HITRAN database for candidate transitions near each detected feature in the 9.7-10.61,\micron\ range that are not contaminated by \ce{H2} or \ce{H2O} emission. Because prompt OH emission arises from non-thermal radiative decay following \ce{H2O} photodissociation, we selected, among the possible candidates, the transition with the highest Einstein $A$ coefficient, which is expected to dominate the observed emission.
The full list of selected transitions for each species is provided in \autoref{tab:selected_lines}.

\begin{deluxetable}{llr}[ht!]
\tablecaption{Wavelengths of transitions used in the stacked-line analysis.}
\tablehead{
\colhead{Species} & \colhead{Line Type} & \colhead{$\lambda$ [$\mu$m]}
}
\startdata
\ce{H2O} (rot.) & Emission & 15.62568 \\
               &          & 15.83495 \\
               &          & 15.96622 \\
               &          & 16.27136 \\
               &          & 16.50525 \\
               &          & 16.54402 \\
               &          & 17.10254 \\
               &          & 17.14148 \\
               &          & 17.19352 \\
               &          & 21.33317 \\
               &          & 22.37473 \\
               &          & 22.99881 \\
               &          & 24.05845 \\
               &          & 24.91403 \\
\hline
\ce{H2O} (ro-vib.) & Absorption & 5.34529 \\
                   &            & 5.64107 \\
                   &            & 6.07545 \\
                   &            & 6.14316 \\
                   &            & 6.18540 \\
                   &            & 6.34443 \\
                   &            & 6.43355 \\
                   &            & 6.49224 \\
                   &            & 6.52896 \\
                   &            & 6.97738 \\
                   &            & 6.99328 \\
                   &            & 7.14692 \\
\hline
\ce{C2H2} (R-branch) & Absorption & 12.92418 \\
                     &            & 12.96336 \\
                     &            & 13.00280 \\
                     &            & 13.04250 \\
                     &            & 13.08246 \\
                     &            & 13.12268 \\
                     &            & 13.16317 \\
                     &            & 13.20393 \\
                     &            & 13.24495 \\
                     &            & 13.28625 \\
                     &            & 13.32781 \\
                     &            & 13.36966 \\
\hline
\ce{CH4} & Absorption & 7.50344 \\
         &            & 7.97862 \\
         &            & 8.22233 \\
\hline
OH & Emission & 9.79069 \\
            &          & 10.06896 \\
           &          & 10.23102 \\
            &          & 10.40986 \\
            &          & 10.60681 \\
\enddata
\end{deluxetable}
\label{tab:selected_lines}

\section{SED Modeling}\label{app:sed_modeling}

Infrared spectral indices are typically used to establish the evolutionary stage of young stellar objects. However, many degeneracies between inclination, foreground extinction, and residual envelope emission can produce similar indices for physically distinct systems, motivating a complementary analysis based on full SED modeling. To explore this, we use a grid of pre-calculated SEDs for young stellar objects. These radiative transfer models, first presented in \citet{robitaille2006}, have been recently updated and now include JWST observations \citep{robitaille2017, richardson2024}, and have been widely used to characterize surveys of YSOs \citep[e.g.][]{benedettini2018,habel2024}. In short, these models are made up of three components: a star, a disk, and an envelope, and they can include contributions from the ambient ISM, significant for line-of-sight extinction. The grid spans a broad range of stellar properties, disk and envelope masses, infall rates, cavity opening angles, and inclinations, and synthetic photometry is pre-computed for a large set of filters. For our fitting, we consider only the models that have a disk, but allow the ambient ISM and envelope to be optional. We use the photometry compiled in the AGE-PRO survey \citep{zhang2025,ruiz-rodriguez2025}, and we calculate photometric fluxes for the MIRI filters from the MRS spectrum. To have a handle on the far-infrared excess, we take a \textit{Herschel}/PACS measurement from \citet{marton2024}. However, as the flux for this observation was extracted $\sim$1.5" away from the center of the source, we take this to be an upper limit on the surrounding envelope emission. During the fitting, we allow for an 80\% confidence on the upper limit, which permits models to marginally exceed the measured flux while penalizing larger deviations through an increased $\chi^2$ (see \citet{robitaille2017zndo} for more details), thereby accounting for uncertainties in the spatial origin of the emission.

\begin{figure*}[ht!]
\includegraphics[width=18cm]{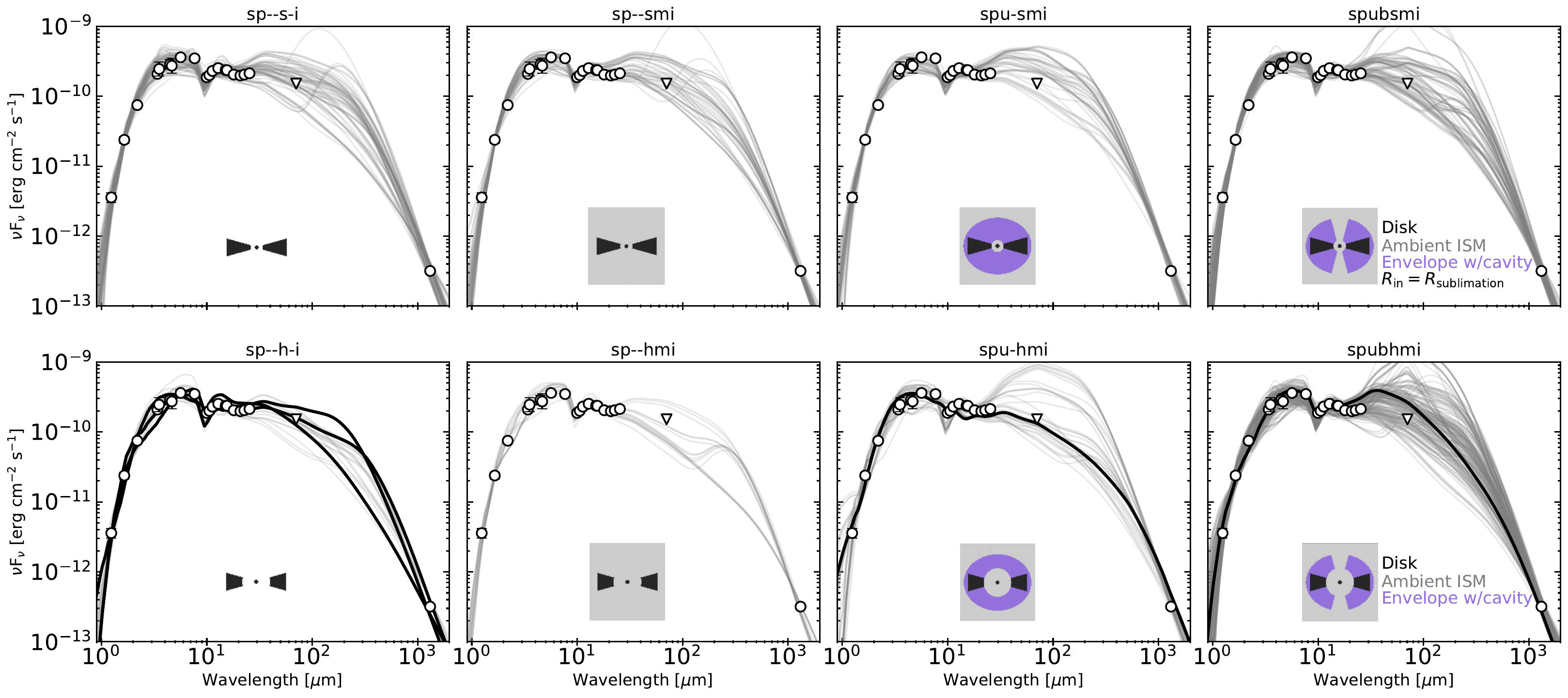}
\caption{Best-fit SED models for ISO-Oph 37. Each panel shows a different model set, with the icon describing the contributions. Details on the specific model names are explained in \citet{robitaille2017}. }
\label{fig:all_seds}
\end{figure*}

We fit ISO-Oph~37 using the \texttt{sedfitter} routine \citep{robitaille2017,robitaille2017zndo} with the updated models from \citet{richardson2024}, which compares the observed SED to this model grid. For each model, \texttt{sedfitter} scales the SED to the allowed distance range, applies a foreground extinction drawn from a specified $A_V$ interval (in our case, $60>A_v>0$), and computes a $\chi^2$ statistic using the observed fluxes and uncertainties. Models that require extreme scaling are rejected, and the remaining models are ranked by reduced $\chi^2$. We show a subset of the fitted models in \autoref{fig:all_seds}. It is important to note that some of these models, even though they reproduce the emission, have parameters that are beyond the credible intervals, e.g. the stellar temperatures or intrinsic source luminosity are too high, the inclination is too low, etc. We emphasize that rather than adopting a single ``best'' model, the goal is to examine the ensemble of acceptable fits to assess whether disk-only configurations suffice or whether an additional envelope component is required, and to explore how inclination and extinction trade off in reproducing the observed SED. Therefore, we further select a subset of models that has stellar temperature, inclination, intrinsic source luminosity and extinction within reasonable estimates of the values in \citet{ruiz-rodriguez2025} and \citet{vioque2025}, corresponding to $T_{\rm eff}=3970$~K, $i=72.59^\circ$, $\log L_\star=-0.06$ and $A_{\rm v}=16.2$, respectively. These resulting models are shown as black lines in \autoref{fig:all_seds}.  

Three of the best-fit models are matched with the \texttt{sp--h-i} models, that have contributions from a star and a passive disk, with the inner radius not fixed at the sublimation radius. The other two models correspond to configurations that include an envelope and ambient ISM. None of the models seem to prefer the configuration where the inner radius is set at the dust sublimation radius. Based on these five ``best-fit" models, we can infer that the observed SED does not favor the absence or presence of an envelope, highlighting the evolutionary ambiguity of this source.

\bibliography{a-main}{}
\bibliographystyle{aasjournalv7}

\end{CJK*}
\end{document}